\documentclass[12pt]{extarticle}
\usepackage[margin=16mm]{geometry}

\usepackage{amsmath,amsthm,amssymb,mathtools,comment,bbm,enumitem}
\usepackage{graphicx,color}
\usepackage{bm,makecell,pict2e,diagbox}
\usepackage[bookmarks=true,bookmarksnumbered=true,bookmarkstype=toc]{hyperref}
\usepackage{mathptmx,lineno}
\usepackage{algorithmicx}
\usepackage{epstopdf,multirow,colortbl}
\usepackage[table,xcdraw]{xcolor}

\def\argmin{\mathop{\rm argmin}}

\theoremstyle{definition}

\makeatletter
\@addtoreset{equation}{section}
\makeatother

\usepackage[most]{tcolorbox}
  \tcbset{
    colback=white,
    colframe=black!10,
    coltitle=black,
    boxsep=1.0ex,
    left=1.2ex,right=1.2ex,top=1.0ex,bottom=1.0ex,
    fonttitle=\bfseries
  }
\tcbuselibrary{breakable}

\title{Stopping criteria in structural reliability analysis: An overview\footnotetext[0]{\noindent 
Email address: raykawai@g.ecc.u-tokyo.ac.jp.
Affiliation: Graduate School of Arts and Sciences / Mathematics and Informatics Center, The University of Tokyo, Japan.
This research was partially supported by JSPS Grants-in-Aid for Scientific Research 21K03347 and 24K06844.}}
\author{\sc Reiichiro Kawai}
\date{}

\begin{document}

\maketitle

\begin{abstract}
\noindent 
Stopping criteria have become a crucial component of probabilistic and statistical methods in structural reliability analysis, helping prevent structural defects and failures, as well as inaccurate evaluations of existing structures.
Over time, these rules have evolved in both concept and application, whereas their treatment in the literature remains fragmented.
They are often developed and presented in connection with specific models or analytical frameworks, obscuring their broader applicability across diverse models and methodologies.
On this basis, the aim of this article is twofold.
First, it overviews a wide range of stopping rules, drawing on concepts from applied probability and statistics.
These strategies include classical methods based on error estimates or variance, and stability detection.
While many of these ideas were developed in other fields, they have the potential to offer useful tools for structural reliability tasks like managing sampling, improving surrogate models, and checking for convergence.
\textcolor{black}{Second, this article introduces a structured framework to organize past and recent developments in this area. 
It brings together various stopping criteria, clarifies their conceptual roles, explains their theoretical foundations, and compares their practical strengths and limitations. 
By relating these methods to common modeling and methodological practices, and by incorporating a systematic comparison of performance aspects such as accuracy, computational cost, and robustness, we aim to identify dominant strategies, highlight shared challenges, and point to directions for further progress.}

\vspace{0.3em}
\end{abstract}

\tableofcontents

\section{Introduction}\label{section introdution}

In structural reliability analysis, stopping criteria -- alternatively referred to as stopping rules, convergence criteria or termination conditions -- define the conditions under which iterative computational procedures terminate, ensuring that reliability estimates are obtained with sufficient accuracy and efficiency.
These criteria are particularly relevant in approximate and simulation methods, where convergence behavior is often nontrivial.
Depending on the approach employed, stopping rules may be built upon the stabilization of probability estimates \cite{SONG2023103479}, checking points and sensitivity measures in reliability index formulations \cite{doi:10.1061/JMCEA3.0001848}, or computational constraints related to available resources.
Various formulations exist, ranging from fixed iteration limits to adaptive thresholds that dynamically respond to numerical behavior \cite{bucher2009computational}.

The selection of appropriate stopping criteria is crucial for balancing computational efficiency and accuracy.
Excessive iterations may incur unnecessary computational costs without significant gains in precision \cite{rubinstein2016simulation}, whereas premature termination can lead to unreliable probability estimates \cite{Melchers_Beck_2018}, potentially resulting in structural defects or failures \cite{REN2024102407} and improper reliability assessment of existing structures \cite{DIAMANTIDIS2025102452}, as a result of human error in specifying appropriate stopping criteria.
Well-defined stopping rules ensure that iterative or simulation-based procedures reach stable and meaningful convergence, preventing both under-processing and over-processing \cite{AU2001263}.
In adaptive methods, such criteria regulate the allocation of computational resources according to the observed convergence behavior, balancing accuracy and efficiency \cite{PAPAIOANNOU201666}.
Conversely, the absence of well-formulated stopping conditions can render reliability evaluations either computationally prohibitive or statistically unreliable, thereby undermining their practical value in engineering contexts.

The development of stopping criteria can date back to early studies on Monte Carlo simulation, where researchers sought practical ways to determine convergence in probabilistic estimations \cite{Ray_1957, Stein_Wald_1947}.
As reliability methods evolved, variance-based stopping rules became prominent, particularly in importance sampling and stratified sampling techniques \cite{SONG2023103479}.
The introduction of subset simulation \cite{AU2001263} refined stopping criteria by incorporating adaptive thresholds based on intermediate failure probabilities.
In parallel, iterative reliability algorithms such as FORM and SORM \cite{doi:10.1061/(ASCE)0733-9399(1984)110:3(357), doi:10.1061/JMCEA3.0001848} necessitated convergence checks based on sensitivity measures and design point stability.
These foundational developments paved the way for modern adaptive stopping rules that enhance computational efficiency while maintaining accuracy in reliability assessments.

Despite extensive research, stopping criteria remain diverse and fragmented, each requiring specific assumptions and conditions depending on the computational method, problem formulation, and accuracy requirements.
This diversity makes systematic comparison challenging and complicates the selection of appropriate criteria for specific applications.
A comprehensive overview is therefore beneficial to consolidate existing knowledge, clarify technical foundations, and provide a structured perspective on the strengths and limitations of various stopping rules.
Such an effort would not only enhance understanding but also facilitate the development of more robust and adaptable stopping criteria in structural reliability analysis.

\subsection{Existing reviews and benchmark studies}
\label{subsection existing reviews and benchmark studies}

To the best of our knowledge, the present article represents the first effort in the literature to systematically address stopping criteria in the context of structural reliability analysis.
In contrast with prior reviews and benchmarking studies on numerical methodologies, this article is organized around a structured framework aimed at guiding the systematic use of stopping criteria.

Although there exist numerous monographs and book chapters that discuss computational methods in structural reliability analysis, a focused and thorough treatment of stopping criteria has generally been lacking. 
Seminal works typically concentrate on broader mathematical modeling and methodological aspects, providing limited explicit coverage of stopping criteria, such as \cite{bucher2009computational, Casciati_Roberts_1996, Ditlevsen_Madsen_1996, Faulin_Juan_Martorell_RamirezMarquez_2010, expectationconstrained, Lemaire_2013, Mahmoodian_2018, Melchers_Beck_2018, Nowak_Collins_2012, ROY2023109126, TABANDEH2022102216, ThoftCristensen_Baker_2012, Wang_2021, Zhang03032024}.

While specialized reviews and benchmark studies occasionally discuss stopping criteria, such considerations have typically not been their primary focus. Instead, stopping criteria often appear briefly within broader topics such as surrogate modeling \cite{ECHARD2011145, TEIXEIRA2021102019}, active learning strategies \cite{MOUSTAPHA2022102174}, artificial neural networks \cite{Chojaczyk_Teixeira_Neves_Cardoso_GuedesSoares_2015}, and broader machine learning frameworks \cite{SaraygordAfshari_2022}.
Recent advancements in these areas  address stopping rules explicitly, particularly through convergence diagnostics, error metrics, and statistical criteria, albeit not as the primary subject of investigation \cite{ELLINGWOOD2025102474, Lee_2022, TEIXEIRA2021102019}.

The present overview seeks to fill this gap by systematically examining the development, implementation, and comparative effectiveness of stopping criteria across a variety of probabilistic modeling and methodological frameworks in structural reliability analysis.
In doing so, we aim to provide foundational insights and offer guidance for future advances in this critical yet underexplored aspect of structural reliability analysis.
In line with our primary objective of providing a comprehensive methodological guide, the emphasis is placed on stopping criteria and closely related techniques, without extending the scope to performance assessments, benchmark comparisons of specific algorithms, or to other areas of structural engineering analysis, such as decision-making under uncertainty \cite{FERESHTEHNEJAD2017113}, robust design optimization \cite{CHAKRABORTY2021103139, LEE2022103218}, experimental design \cite{DECARVALHODANTASMAIA2025103724, SHI2024103551}, optimal stopping for software rejuvenation \cite{MACHIDA2017128}, condition-based maintenance optimization \cite{ROSQVIST2002319}, or life-cycle cost optimization \cite{GOMES201448, TAFLANIDIS2009508}, although these domains also employ stopping criteria in one form or another.

\color{black}
\subsection{Taxonomy}
\label{subsection taxonomy of stopping criteria}

To reduce fragmentation and enhance analytical consistency, this review adopts a unified, mechanism-based taxonomy of stopping criteria, which serves as the primary analytical lens throughout the article.
Rather than classifying rules by modeling or simulation frameworks, we organize them according to their underlying decision mechanisms.
Specifically, as summarized in Table \ref{tab:taxonomy}, five categories are identified: (i) variance-based criteria, which rely on measures of statistical uncertainty (for instance, coefficient of variation); (ii) error-based criteria, defined by prescribed absolute or relative accuracy; (iii) stability-based criteria, which track convergence through successive estimate changes; (iv) budget-based criteria, which impose limits on computational effort; and (v) confidence- or classification-based criteria, commonly employed in surrogate and active learning frameworks to quantify predictive uncertainty or misclassification risk.

\begin{table}[!ht]\color{black}
\centering
\caption{Mechanism-oriented taxonomy of stopping criteria.}

\vspace{0.5em}
\begin{tabular}{p{0.18\columnwidth} | p{0.35\columnwidth} | p{0.35\columnwidth}}
\hline
\textbf{Category} & \textbf{Principle} & \textbf{Typical indicators} \\ 
\hline
Variance-based 
& Terminate when statistical uncertainty falls below a prescribed level 
& Coefficient of variation (CoV), variance estimates \\ \hline

Error-based 
& Ensure estimated quantity satisfies a predefined accuracy requirement 
& Relative/absolute error bounds, $(\varepsilon,\delta)$-criteria \\ \hline

Stability-based 
& Monitor convergence through successive estimate changes 
& $|P_f(n+1)-P_f(n)|$, moving averages \\ \hline

Budget-based 
& Impose limits on computational resources or sampling effort 
& Maximum number of samples, time or cost constraints \\ \hline

Confidence- or classification-based 
& Assess predictive certainty or classification reliability (often in surrogate models) 
& Misclassification probability, confidence intervals, learning function thresholds \\
\hline
\end{tabular}
\label{tab:taxonomy}
\end{table}

This mechanism-oriented taxonomy provides a unified framework for interpreting stopping rules across otherwise heterogeneous methodologies.
In the remainder of the article, these categories serve as the primary analytical lens, while specific approaches, such as approximate reliability methods, Monte Carlo simulation, advanced sampling techniques, and surrogate-based models, are treated as application domains in which different classes of stopping criteria are instantiated and combined.
This structure promotes cumulative understanding and reduces overlap by linking diverse approaches through their shared underlying principles.

To further clarify the conceptual scope of this review, it is essential to distinguish between closely related notions concepts that are sometimes used interchangeably in the literature. 
We define a ``stopping criterion'' as an explicit rule that determines when an algorithm or simulation is terminated. 
In contrast, a ``selection rule'' governs the choice of the next sample or evaluation point (for instance, learning functions in adaptive sampling), while a ``convergence diagnostic'' provides an indicator of progress without necessarily prescribing termination. 
Although these notions are conceptually distinct, many practical frameworks combine them, where stopping decisions are often informed by diagnostics or driven indirectly by selection mechanisms. 
In this review, we adopt a broad perspective while aiming to clearly indicate the role each quantity plays in the stopping process.

\subsection{Performance dimensions}
\label{subsection performance dimensions}

In addition to categorizing stopping criteria by mechanism, it is important to evaluate them according to consistent performance dimensions. 
In Table \ref{table:performance_metrics}, we summarize the main metrics commonly used in the literature.
Across the literature, stopping criteria are commonly assessed in terms of their impact on failure probability bias, variance or coefficient of variation, computational cost (for instance, number of model evaluations), robustness to rare events, and sensitivity to model or surrogate error. 
These dimensions provide a basis for comparing different stopping strategies beyond their formal definitions and are used throughout this review to interpret their practical effectiveness.

\begin{table}[!ht]\color{black}
\centering
\caption{Common performance metrics used to evaluate stopping criteria.}
\vspace{0.5em}

\begin{tabular}{p{0.2\columnwidth} p{0.32\columnwidth} p{0.38\columnwidth}}
\hline
\textbf{Metric} & \textbf{What it measures} & \textbf{Typical interpretation / limitations} \\
\hline

Failure probability bias 
& Deviation between estimated and true failure probability 
& Direct indicator of accuracy; true value often unknown in practice, requiring surrogate benchmarks or reference solutions \\

\hline

Coefficient of variation (CoV) 
& Relative statistical dispersion of the estimator 
& Widely used measure of uncertainty; may be unreliable in rare-event settings due to insufficient failure samples \\

\hline

Confidence interval width 
& Statistical uncertainty in probability estimates 
& Provides probabilistic bounds on estimation; depends on distributional assumptions and may be overly conservative \\

\hline

Number of model evaluations 
& Computational cost of simulation or learning process 
& Key efficiency measure; does not directly reflect accuracy or robustness \\

\hline

Convergence/stability indicators 
& Change in estimates across iterations 
& Easy to implement; may indicate local convergence without ensuring global correctness \\

\hline

Robustness to rare events 
& Sensitivity to small failure probabilities or complex domains 
& Critical in reliability analysis; difficult to quantify consistently across studies \\

\hline

Sensitivity to model/surrogate error 
& Impact of approximation error on stopping decisions 
& Important in surrogate-based methods; often problem-dependent and rarely reported explicitly \\

\hline

\end{tabular}
\label{table:performance_metrics}
\end{table}

A central issue that emerges from the present review concerns the status of benchmarking and reproducibility in the evaluation of stopping criteria.
While widely recognized as important, benchmarking in this domain is not only limited but structurally fragmented.
Existing studies assess stopping behavior using heterogeneous performance metrics (for instance, failure probability bias, coefficient of variation, computational cost), under non-comparable modeling assumptions and problem settings, particularly in rare-event regimes.
As a result, direct cross-method comparison is often ambiguous and, in some cases, conceptually ill-posed. 
Rather than attempting to impose a unified benchmark framework, this review adopts a structured evaluation perspective that organizes these disparate practices along a small set of core performance dimensions, thereby clarifying both the scope and the inherent limitations of current comparative evidence.

\subsection{Review methodology}
\label{subsection review methodology}

Building upon the taxonomy (Section \ref{subsection taxonomy of stopping criteria}) and the defined performance dimensions (Section \ref{subsection performance dimensions}), this section outlines the review methodology employed to identify, select, and organize the literature within this framework.
To ensure transparency and reproducibility, this review adopts a structured, though not fully systematic, approach to gathering and synthesizing the relevant literature on stopping criteria in structural reliability analysis.

\paragraph{Literature search strategy:}

Relevant publications have been identified through major scientific databases, including Web of Science, Scopus, and Google Scholar.
The search focuses on combinations of keywords such as ``stopping criteria,'' ``convergence,'' ``termination condition,'' ``structural reliability,'' ``Monte Carlo simulation,'' ``rare-event simulation,'' ``active learning,'' and ``surrogate modeling.''
Both backward and forward citation tracking have been employed to expand the corpus and to ensure coverage of influential and frequently cited works.

\paragraph{Scope and criteria:}

The review targets contributions that explicitly propose, analyze, or apply stopping criteria within the context of structural reliability analysis or closely related domains (for instance, uncertainty quantification and reliability-based design).
Priority is given to peer-reviewed journal articles, well-established conference proceedings, and, where appropriate, publicly available preprints.
Foundational works in probability and statistics are also included when they underpin commonly used stopping rules (for instance, variance-based or error-based criteria).
Publications are included if they provide either a formal stopping rule, an implicit termination strategy embedded in an algorithm, or a critical discussion of convergence behavior relevant to reliability estimation.



\paragraph{Classification and organization:}

To address the fragmented nature of the literature, the collected works are organized according to the unified taxonomy of stopping criteria based on their underlying mechanisms, as introduced in Section \ref{subsection taxonomy of stopping criteria}.
Specifically, stopping rules are categorized into classes such as variance-based, error-based, stability-based, budget-based, and confidence-based criteria. 
In parallel, the analysis is guided by the performance dimensions defined in Section \ref{subsection performance dimensions}, which provide a complementary perspective for evaluating and comparing these criteria in terms of efficiency, accuracy, robustness, and computational cost. 
The integration of mechanism-oriented classification with performance-based evaluation establishes a consistent analytical lens across diverse modeling frameworks and simulation strategies. 
Subsequent sections examine how these categories and performance considerations manifest within specific methodological contexts, including approximate reliability methods, Monte Carlo simulation, advanced sampling techniques, and surrogate-based approaches.

\paragraph{Limitations:}

Despite efforts to ensure breadth and balance, this review does not claim to be exhaustive.
The selection process inevitably involves subjective judgment, particularly in emerging areas where stopping criteria are often embedded implicitly within broader algorithms.
Nevertheless, by making the search strategy, inclusion criteria, and classification principles explicit, the review aims to provide a transparent and structured synthesis that supports both interpretation and future extension.

\color{black}

\color{black}

\subsection{Structure of the present overview}
\label{subsection structure of the present overview}

This article is set out as follows.
Following the introduction of general notation and technical preliminaries in Section \ref{subsection preliminaries}, which serve as the foundation for this overview, Section \ref{section approximate methods} turns to stopping criteria for approximate methods. 
While rooted in more classical rather than modern frameworks, these approximate methods constitute fundamental building blocks for many of the reliability and computational approaches surveyed in following sections.

In Section \ref{section stopping criteria for monte carlo simulation} beginning with the basic principles (Section \ref{subsection basics of monte carlo simulation}), we review stopping rules based on the coefficient of variation (Section \ref{subsection CoV}), the so-called principled Monte Carlo simulation (Section \ref{subsection error-based stopping criteria}), and absolute error-based criteria (Section \ref{subsection absolute error MC}), the latter framed within applied probability through Chebyshev's inequality (Section \ref{subsubsection chebyshev}), the central limit theorem (Section \ref{subsubsection CLT}), Hoeffding's inequality (Section \ref{subsubsection hoeffding}), and stability-based stopping criteria (Section \ref{subsection stability-based stopping criteria}), along with summarizing remarks (Section \ref{subsection summary 3}).
These have been developed primarily in applied probability and statistics, underscoring their relevance and potential to advance stopping strategies in structural reliability analysis.
For readers mainly interested in existing applications, it may suffice to skip much of these subsections and proceed directly to the pertinent topics in Sections \ref{section sampling methods}, \ref{section advanced modeling frameworks}, and \ref{section advanced methodological frameworks}.

Next, Section \ref{section sampling methods} addresses stopping criteria for advanced sampling methods, such as importance sampling, subset simulation, line sampling, directional sampling and Markov chain Monte Carlo, as presented in Sections \ref{subsection importance sampling}-\ref{subsection MCMC}.

In Sections \ref{section advanced modeling frameworks} and \ref{section advanced methodological frameworks}, we aim to highlight the overarching landscape of stopping criteria across various probabilistic and statistical modeling and methodological frameworks.
Instead of focusing on the fine-grained technicalities of specific models or methods, we emphasize a higher-level perspective that maintains a coherent flow and helps the reader discern the general patterns and tendencies.

First, in Section \ref{section advanced modeling frameworks}, we discuss stopping criteria around key advanced modeling frameworks, such as Kriging models,
response surfaces,
polynomial chaos expansions,
support vector machines,
neural networks,
multifidelity models,
reduced-order models,
and ensemble models, 
as presented respectively in Sections \ref{subsection kriging models}-\ref{subsection ensemble models}, along with summarizing remarks (Section \ref{subsection summary 5}).

Building on this modeling-oriented perspective, next in Section \ref{section advanced methodological frameworks}, we reorganize the discussion in the context of advanced methodological approaches.
These comprise active learning reliability methods,
Bayesian methods,
adaptive response surface methods,
reliability-based design optimization,
and time-variant reliability analysis,
as presented in Sections \ref{subsection active learning}-\ref{subsection time-variant reliability analysis}.
Let us stress that many references can recur throughout Sections \ref{section sampling methods}, \ref{section advanced modeling frameworks} and \ref{section advanced methodological frameworks}, reflecting the frequent use of multiple advanced models and methods in combination.

In conclusion, Section \ref{section concluding remarks} provides both a synthesis of the key discussions and an exploration of prospective research directions, as outlined in Sections \ref{subsection synthesis and critical reflections} and \ref{subsection prospects for future research}.

\color{black}
\subsection{Technical preliminaries}\label{subsection preliminaries}

We first introduce the notation adopted throughout this overview. 
The indicator of an event $E$ is written as $\mathbbm{1}(E)$, while for a set $D$ and a point ${\bf x}$ we use $\mathbbm{1}_D({\bf x})$. 
Given two vectors ${\bf x}$ and ${\bf y}$ of the same dimension, their inner product is denoted by $\langle {\bf x}, {\bf y} \rangle = {\bf x}^{\top}{\bf y}$. 
The symbol $\nabla$ stands for the gradient operator. 
Finally, the $(d-1)$-dimensional unit sphere in $\mathbb{R}^d$ is denoted by $S^{d-1} := \{ {\bf v}\in\mathbb{R}^d : \|{\bf v}\| = 1 \}$.
	
We remark that the symbol $\Omega$ is used here to denote the support of the random vector under consideration, not the sample space in the conventional sense. 
Convergence in distribution is expressed by the notation $\overset{\mathcal{L}}{\to}$.
The standard normal probability density function and its cumulative distribution function are denoted by $\phi(x) := \exp(-x^2/2)/\sqrt{2\pi}$ and $\Phi(x) := \int_{-\infty}^x \phi(y)\,dy$, respectively.
Finally, the gamma function is written as $\Gamma(s) := \int_0^{\infty} x^{s-1} e^{-x}\, dx$ for $s>0$. 

The failure probability can be written in general form as the multidimensional integral
\begin{equation}\label{original formulation}
	P_f := \mathbb{P}\!\left(g(X)\leq 0\right)
	= \int_{\Omega} \mathbbm{1}\!\left(g({\bf x})\leq 0\right) f({\bf x}) \, d{\bf x},
\end{equation}
where $X=(X_1,\cdots,X_d)$ is a random vector defined on the domain $\Omega \subseteq \mathbb{R}^d$, representing the uncertain variables of the structural system of interest. 
The function $f({\bf x})$ denotes the probability density of $X$. 
The mapping $g({\bf x})$ appearing in \eqref{original formulation} is referred to as the limit state function (primarily when emphasizing the boundary between performance states) or as the performance function (when focusing on how well the system performs).
The subset $\{{\bf x}\in \Omega:\, g({\bf x})\leq 0\}$ represents the failure domain, while its boundary $\{{\bf x}\in \Omega:\, g({\bf x})=0\}$ is called the limit state surface, which separates the failure region from the safe region.
In many cases, one introduces a mapping $Z = T(X)$ (commonly termed the isoprobabilistic transformation) is introduced, where $T:\mathbb{R}^d \to \mathbb{R}^d$ is a suitable mapping, provided it is well defined. 
This allows the original failure probability \eqref{original formulation} to be reformulated based upon standard Gaussian random vectors. 
Given an inverse mapping $T^{-1}$, one can define the transformed limit state function $g_T({\bf z}) := g(T^{-1}({\bf z}))$, which is expressed on the standard Gaussian space. 
Accordingly, the failure probability \eqref{original formulation} can be written as
\begin{equation}\label{reliability gaussian formulation}
	P_f
	=\int_{\Omega}\mathbbm{1}\left(g({\bf x})\le 0\right)f({\bf x})d{\bf x}=\int_{\mathbb{R}^d}\mathbbm{1}\left(g_T({\bf z})\le 0\right)\prod_{k=1}^d \phi(z_k)d{\bf z}.
\end{equation}

While not directly addressing stopping criteria, we refer the reader to \cite[Section 5.6.3]{Melchers_Beck_2018} for an interesting discussion on reducing the number of failure modes by truncating insignificant subregions along with some relevant criteria, for instance, when the failure region $\{{\bf x}\in \Omega:\,g({\bf x})\le 0\}$ can be decomposed into several subregions as the union $\cup_{k=1}^n \{{\bf x}\in \Omega:\,g_k({\bf x})\le 0\}$.

\section{Stopping criteria for approximate methods}
\label{section approximate methods}

\textcolor{black}{Approximate approaches, such as first- and second-order reliability methods and certain higher-moment techniques, offer computational efficiency in structural reliability analysis, while their accuracy in real-world applications is often limited \cite{10.1098/rsta.2022.0395, YANG2026111916, YANG2025459}.}
We overview their fundamental principles and a few representative stopping rules, recognizing that, despite their limitations, these methods constitute essential building blocks for numerous reliability and simulation frameworks discussed later.

Recalling the transformed limit state function $g_T({\bf z}):=g(T^{-1}({\bf z}))$, the main idea of the first-order reliability method (FORM) \cite{doi:10.1061/JMCEA3.0001848} is to approximate the failure domain $\{{\bf z}\in \mathbb{R}^d:\,g_T({\bf z})\le 0\}$ by a half-space fitted to the true failure domain at the point of the maximum probability density, with reference to a linear approximation for the limit state function in the formulation \eqref{reliability gaussian formulation} through a first-order Taylor series expansion of the limit state function about the point ${\bf z}^*$, as $g_T({\bf z})\approx g_T({\bf z}^*)+\langle {\bf z}-{\bf z}^*, \nabla g_T({\bf z}^*)\rangle$.
The point ${\bf z}^*$ is termed the checking point (also known as the most probable point or the design point). 
It is defined as a point on the limit state surface $\{{\bf z}\in\mathbb{R}^d : g_T({\bf z})=0\}$, having the minimal Euclidean distance from the origin. 
This distance, known as the Hasofer-Lind reliability index (or, simply, the reliability index), is defined as $\beta := \|{\bf z}^*\|$, with ${\bf z}^* := \argmin_{g_T({\bf z})=0}\|{\bf z}\|$.
Once the checking point ${\bf z}^*$ is identified, the first-order reliability method (FORM) approximates the limit state surface by the tangent hyperplane at ${\bf z}^*$. 
Under the assumption that the tail probability of the standard normal distribution provides a good approximation to the probability content of the failure region $\{{\bf z}\in\mathbb{R}^d : g_T({\bf z}) \leq 0\}$, the failure probability is well approximated in the way of $P_f\approx \Phi(-\beta)=\Phi(-\|{\bf z}^*\|)$.

The accuracy of the FORM approximation is however ensured in the limited case where the limit state function is nearly linear, while the function exhibits substantial nonlinearity in most applications.
To address this issue, the second-order reliability method (SORM) \cite{Fiessler_Rackwitz_Neumann_1979} enhances the approximation using a second-order Taylor expansion, so that the failure surface is approximated by a paraboloid at the checking point, leading to improved estimates when curvature effects are significant.

The main computational task is hence to locate a checking point through a suitable search algorithm, such as the Hasofer-Lind-Rackwitz-Fiessler (HL-RF) algorithm \cite{doi:10.1061/JMCEA3.0001848, RACKWITZ1978489}:
\[
 {\bf z}_{k+1}=\frac{(\nabla g_T({\bf z}_k))^{\top} {\bf z}_k-g_T({\bf z}_k)}{\|\nabla g_T({\bf z}_k)\|^2}\nabla g_T({\bf z}_k),
\]
and a variety of its modified and improved versions.
Clearly, the search may become computationally demanding, especially when evaluating the limit state function, its gradient, and associated quantities is nontrivial.

In sharp contrast to the substantial body of research on convergence analysis, the topic of stopping criteria has largely drawn upon established results from nonlinear optimization. For example, when the original sources of randomness (that is, $X$ and ${\bf x}$, rather than $Z$ and ${\bf z}$, in the formulation \eqref{reliability gaussian formulation}) are bounded, the HL-RF algorithm is examined in \cite{BECK201679} with three stopping criteria: the proximity of the limit state function to zero ($|g_T({\bf z}_k)|<c$),  the satisfaction of the first-order optimality condition, both in the standard Gaussian space, and the smallness of the update norm in the original space (such as $\|{\bf x}_{k+1}-{\bf x}_k\|<\epsilon$, for some small tolerance $\epsilon>0$).
In turn, stopping criteria in software implementations are often defined in an ad hoc manner.
For instance, PHIMECA-SOFT \cite[Section 2.2.4]{LEMAIRE2006130} employs the conditions $g_T({\bf z}_{n+1})/g_T({\bf z}_0)\le \epsilon_G$ and $(\beta_{n+1}-\beta_n)/\beta_n\le \epsilon_{\beta}$ to terminate the search, where $\epsilon_G$ and $\epsilon_{\beta}$ are tolerances specified by the user.

In particular, the latter criterion $(\beta_{n+1}-\beta_n)/\beta_n\le \epsilon_{\beta}$ and its variants, often referred to as the beta stability, have been adopted as a key component of stopping criteria in various modern frameworks.
In contrast, many stopping criteria have also been built upon the so-called beta bounds, often quantifying the approximation error $|\beta^+-\beta^-|$ or, in relative form $|\beta^+-\beta^-|/\beta$, based on certain upper and lower bounds $\beta^+$ and $\beta^-$ in estimating the stability index $\beta$ (for instance, \cite{MOUSTAPHA2022102174}).
Those $\beta$-based stopping criteria have been adopted in a wide variety of sampling, advanced modeling and methodological frameworks.

An important class of extensions builds upon these core ideas to better capture the behavior of complex or nonlinear systems. For example, the evolutionary tail-equivalent linearization (ETEL) method \cite{BROCCARDO2021102044} provides a nonlinear stochastic dynamic analysis framework by iteratively linearizing the system response in the tails of the distribution. This approach, while improving accuracy in the estimation of rare event probabilities, requires carefully designed stopping criteria to avoid unnecessary iterations and to ensure computational tractability in dynamic reliability analyses.
We note that the reliability index serves as a key component in variance-based and stability-based stopping criteria (Section \ref{section approximate methods}) used in comprehensive benchmark analyses of active learning reliability approaches \cite{MOUSTAPHA2022102174}.

Another major challenge is the propagation of uncertainties characterized by probability boxes (p-boxes) or hybrid probabilistic models.
The hybrid method proposed in \cite{LIU201773} combines classical probabilistic analysis with p-box representations, allowing for the quantification of both aleatory and epistemic uncertainties.
In this context, stopping criteria must be adapted to ensure convergence not only in the classical sense (for instance, changes in the reliability index or the design point) but also in terms of the stabilization of the bounds of failure probabilities as both probability and interval uncertainties are resolved through iterative procedures.

Similarly, computational methods for propagating structural uncertainties, such as those described in \cite{CAMBIER200229}, demonstrate the importance of stopping criteria in the context of dynamic structural analysis.
Here, iterative algorithms are employed to account for uncertainties in dynamic behavior prediction, with termination often guided by thresholds on residuals, changes in dynamic response quantities, or the convergence of system reliability measures.

Although the methodologies range widely, from iterative linearization to hybrid probabilistic propagation and dynamic response modeling, those all share the central objective of balancing convergence precision with computational efficiency in distinct manners.
Since approximate methods underpin many of modern advanced numerical schemes rather than standing apart from them, various corresponding stopping criteria naturally reappear in the subsequent sections.

\section{Stopping criteria for Monte Carlo simulation}
\label{section stopping criteria for monte carlo simulation}

Formulated as the integrals in \eqref{original formulation} or \eqref{reliability gaussian formulation}, the failure probability may be evaluated either analytically or numerically.
Exact solutions are feasible only in rare, highly idealized situations with limited practical relevance.
In contrast, Monte Carlo simulation (MCS) has long served as the principal tool for numerical approximation in structural reliability analysis.

Here, we review the development and assessment of stopping criteria for Monte Carlo simulation.
\textcolor{black}{In contrast to selection-driven approaches (such as those based on learning functions), the criteria discussed in this section are explicit termination rules, defining when the simulation is stopped once a prescribed level of accuracy or stability is achieved.}
Following a brief overview of its essential principles (Section \ref{subsection basics of monte carlo simulation}), we discuss two widely used frameworks: coefficients of variation (Section \ref{subsection CoV}) and error-based stopping rules (Section \ref{subsection error-based stopping criteria}).
We also examine absolute error-based criteria (Section \ref{subsection absolute error MC}) and stability-based criteria (Section \ref{subsection stability-based stopping criteria}), which, despite their relevance, have not yet been systematically explored in structural reliability analysis.
Originating largely from applied probability and statistics (for instance, \cite{handbook, expectationconstrained, doi:10.1287/mnsc.28.5.550, survey_SteadyStateI, JWsurvey}), these approaches hold promise for advancing stopping strategies in future research.
The reader primarily interested in practical applications may therefore skip much of the later subsections and proceed directly to the topics in Sections \ref{section sampling methods}, \ref{section advanced modeling frameworks}, and \ref{section advanced methodological frameworks}, as appropriate.

\textcolor{black}{While the stopping criteria discussed in this section originate primarily from probability theory and statistical simulation, their practical relevance in structural reliability analysis varies significantly.
In particular, some criteria are directly adopted in Monte Carlo-based reliability estimation, whereas others remain largely theoretical or require modification in rare-event settings. 
In the following, we therefore highlight not only their mathematical properties, but also their applicability and limitations in structural reliability analysis.
These observations also motivate the development of more advanced stopping strategies discussed in later sections, particularly in surrogate-based and adaptive learning frameworks (Sections \ref{section advanced modeling frameworks} and \ref{section advanced methodological frameworks}).}

\subsection{The basics of Monte Carlo simulation}
\label{subsection basics of monte carlo simulation}

The principle of Monte Carlo simulation is to approximate the integral \eqref{original formulation} by repeatedly evaluating the limit state function $g({\bf x})$ on many random samples.
The (crude) unbiased estimator is defined as follows:
\begin{equation}\label{MC estimation}
		P_f(n):=\frac{1}{n}\sum_{k=1}^n\mathbbm{1}\left(g(X_k)\leq 0\right),\quad n\in\mathbb{N},
\end{equation}
where $\{X_k\}_{k\in\mathbb{N}}$ are independent random vectors, identically distributed in accordance with the probability density function $f({\bf x})$ in the integral formulation \eqref{original formulation}.
We have $\mathbb{E}[P_f(n)]=P_f$ for all $n\in\mathbb{N}$, that is, the estimator $P_f(n)$ exhibits unbiasedness at every iteration.
Also, the estimator has decreasing variance with $n$, according to ${\rm Var}(P_f(n))=P_f(1-P_f)/n$.
It also converges to the failure probability $P_f$ by the strong law of large numbers: 
	\begin{equation}\label{SLLN}
	P_f(n)\to P_f,\quad n\to +\infty,
	\end{equation}
with probability one.
Moreover, by the central limit theorem, we have
	\begin{equation}\label{CLT0}
		\sqrt{n}\left(P_f(n)-P_f\right)\xrightarrow{\mathcal{L}} \mathcal{N}(0,P_f(1-P_f)), \quad n\to+\infty,
	\end{equation}
due to ${\rm Var}(\mathbbm{1}(g(X_1)\le 0))=P_f(1-P_f)$, as the distribution of $\mathbbm{1}(g(X_1)\le 0)$ is Bernoulli with probability of success given by $P_f(=\mathbb{P}(\mathbbm{1}(g(X_1)\le 0)))$.
The central limit theorem \eqref{CLT0} can be made more practical as follows:
\begin{equation}\label{CLT1}
 \sqrt{n}\frac{P_f(n)-P_f}{\sigma(n)}\xrightarrow{\mathcal{L}} \mathcal{N}(0,1), \quad n\to+\infty,
\end{equation}
where we have denoted by $\sigma(n)$ the empirical standard deviation based on the first $n$ realizations, which is the square root of the empirical variance $\sigma^2(n)$, defined as
\begin{equation}\label{empirical variance}
 \sigma^2(n):=\frac{1}{n-1}\sum_{k=1}^n \left(\mathbbm{1}(g(X_k)\le 0)-P_f(n)\right)^2,\quad n\in\mathbb{N}.
\end{equation}
The primary advantage of Monte Carlo simulation is its convergence rate of $1/\sqrt{n}$, as given in the central limit theorem \eqref{CLT0}, which depends only on the number of samples $n$ and is independent of the problem dimension $d$, in sharp contrast to many other numerical procedures \cite{Schueller_2009, SONG2023103479}.
The central limit theorem above holds even when the independent random variables $\{\mathbbm{1}(g(X_k)\le 0)\}_{k\in \mathbb{N}}$ are not identically distributed if the Lindeberg condition is satisfied \cite{WANG201933}, indicating that the contribution of any individual random variables to the empirical variance is arbitrarily small.

The straightforward implementation of the estimator \eqref{MC estimation} guarantees convergence through the strong law of large numbers \eqref{SLLN}, yet requiring substantial computational cost, especially when rare-event probabilities are involved.
The primary sources of such cost are: (i) the effort required to generate random vectors under complex or high-dimensional distributions, (ii) the repeated and expensive calls to the limit state function, and (iii) the inherently slow approximation dictated by the central limit theorem \eqref{CLT0} and \eqref{CLT1}, which requires many realizations, especially when the failure region has minimal probability mass.
Consequently, simulations must be terminated carefully to balance accuracy with computational effort, an essential consideration for resource-limited or costly simulations \cite[Sections 4]{ELLINGWOOD2025102474}.
Common stopping criteria include a fixed number of iterations or computation time, though more sophisticated alternatives are discussed in what follows.


\subsection{Coefficients of variation}
\label{subsection CoV}

One of the earliest adopted stopping criteria is based on the coefficient of variation (CoV), which can be defined as the ratio of the standard deviation to the mean on the unbiased estimator \eqref{MC estimation}, that is,
\[
\frac{\sqrt{{\rm Var}(P_f(n))}}{\mathbb{E}[P_f(n)]}
=\sqrt{\frac{1-P_f}{nP_f}}.
\]
This quantity behaves like $(n P_f)^{-1/2}$ when the target probability $P_f$ is very small.
Consequently, if the order of the failure probability is $10^{-m}$ for some positive $m$ (typically, $m\in \{2,\cdots,9\}$), achieving a CoV of 10\% requires the sample size $n$ on the order of $10^{m + 2}$.
We refer the reader to \cite{GIOVANIS2022103293, SHIELDS2021106431} for refinements in estimating the coefficients of variation in the context of subset simulation (Section \ref{subsection subset simulation}).

In the context of Monte Carlo simulation in structural reliability analysis \cite{SONG2023103479}, 
the sampling procedure is often designed to terminate when the coefficient of variation reaches a specified threshold (such as $0.05$),
with the number of iterations and the empirical variance subsequently reported.
The unitless nature of the coefficient of variation makes it particularly useful for comparing datasets with differing units or substantially different mean values.
It, however, exhibits sensitivity to fluctuations in the mean, especially when the mean is small.

Apart from their central role in evaluating the design value of resistance \cite{CASTALDO2019136, NOVAK2021102101}, the CoV have found applications as a stopping criterion across various frameworks in structural reliability analysis, not only for MCS but also for various related sampling methods (Section \ref{section sampling methods}), where sampling is terminated when the CoV reaches a specified threshold \cite{LEMAIRE2006130}, such as $0.05$ (\cite{PAPAIOANNOU201666, Schueller_2009, ASDIS, ReliabilitySS, THAPA2022103351} among others), with the number of iterations and the empirical variance subsequently reported.
Even in advanced modeling and methodological frameworks (Sections \ref{section advanced modeling frameworks} and \ref{section advanced methodological frameworks}), the CoV has frequently been adopted as a stopping criterion.

\textcolor{black}{
This criterion is widely adopted in structural reliability analysis, particularly in Monte Carlo and subset simulation methods, due to its direct connection to estimator variance.
However, its effectiveness often deteriorates in rare-event regimes, where variance estimates become unstable and may significantly underestimate uncertainty.
}

\subsection{Error-based stopping criteria}
\label{subsection error-based stopping criteria}

In the context of structural reliability analysis, the term ``error-based stopping'' \cite{wang_ahafieezadeh_2019} (as well as ``the principled Monte Carlo simulation'' \cite[Section 4]{ELLINGWOOD2025102474}) is commonly given to the $(\epsilon,\delta)$-approximation of the failure probability \cite{doi:10.1137/S0097539797315306}, which is defined as a random estimate $P_f(n)$ satisfying 
\begin{equation}\label{relative error} 
 \mathbb{P}\left(\frac{
P_f(n)-P_f
}{P_f}>\epsilon\right)\le \delta, 
\end{equation} 
ensuring that the estimate $P_f(n)$ lies within the range $[(1-\epsilon)P_f,\,(1+\epsilon)P_f]$ with probability $1-\delta$ at least. 
In what follows, we call the quantities $\epsilon$, $1-\delta$, $\delta$, $\mathbb{P}(
P_f(n)-P_f
/P_f >\epsilon)$ (or, $\mathbb{P}(
P_f(n)-P_f
 >\epsilon)$) and $\mathbb{P}(
P_f(n)-P_f
/P_f \le \epsilon)$ (or, $\mathbb{P}(
P_f(n)-P_f
 >\epsilon)$), respectively, the error tolerance, the confidence and significance levels, and the error and coverage probabilities. 

Here, the error probability $\mathbb{P}(
P_f(n)-P_f
/P_f>\epsilon)$ cannot be evaluated exactly, because the distribution of the estimator $P_f(n)$ is generally unknown, as well as because the failure probability $P_f$ to be estimated is certainly unknown. 
A typical approach is then to apply the Chebyshev inequality 
\[
\mathbb{P}\left(\frac{
P_f(n)-P_f
}{P_f}>\epsilon\right)\le \frac{1/P_f-1}{n\epsilon^2},\quad n\in \mathbb{N}, 
\]
which yields the desired criterion $\mathbb{P}(
P_f(n)-P_f
/P_f> \epsilon)\le \delta$ for all $n\ge \lceil (1/P_f-1)/(\delta \epsilon^2)\rceil$. 

Along this line, the work \cite{PAREDES2019106472} suggests the experiment as follows. 
Construct a single experiment $P_f(n)$ with $n=\lceil c_0(1/P_f-1)\epsilon^{-2}\rceil$, repeat this experiment $r=\lceil c_1\ln (1/\delta)\rceil$ times, and return the median of those $r$ empirical means, with $c_0$ and $c_1$ are positive constants, depending on each other. 
Then, the median resides in the desired range $[(1-\epsilon)P_f,(1+\epsilon)P_f]$ with at least probability $1-\delta$. 
The key issues here are that the constants $c_0$ and $c_1$ are not explicit all the time and, more importantly, that the constant $P_f$ in the number of sample $n$ is not available and is exactly what to be estimated. 
As pointed out in \cite{Jiang2014}, this problem remains an open challenge, as this would require a lower bound on the unknown value of the failure probability $P_f$. 

\textcolor{black}{In practice, this difficulty is often addressed through heuristic remedies, such as two-stage procedures or the use of pilot runs to estimate a lower bound for $P_f$, as already described.
However, these approaches introduce an additional layer of approximation and implicitly relax the original theoretical guarantees, since the prescribed sample size is no longer based on verified prior knowledge of the failure probability $P_f$.
Consequently, the resulting stopping rules may only satisfy the $(\epsilon,\delta)$-criterion conditionally or approximately, rather than in a strict non-asymptotic sense. This gap between theory and implementation is rarely made explicit, yet it directly affects the interpretation of probabilistic guarantees in practical applications.
This limitation is particularly critical in the rare-event regime, where the failure probability $P_f$ may be of order $10^{-6}$ or smaller, making the required sample size highly sensitive to an unknown and potentially vanishing quantity.} 

In the context of relative-precision stopping rules, an algorithm is developed in \cite{https://doi-org.utokyo.idm.oclc.org/10.1002/rsa.20654}, with an upper bound for the expected run length by $2\epsilon^{-2}p^{-1}\ln(2\delta^{-1})(1-(4/3)\epsilon)^{-1}$ for general success probability $p\in (0,1)$, and by $(1/5)\epsilon^{-2}p^{-1}(1+2\epsilon)(1-\delta)\ln((2-\delta)/\delta)$ if restricted to $p\in (0,1/2]$. 
This approach improves upon the earlier algorithm of \cite{doi:10.1137/S0097539797315306}, offering advantages such as tighter constants and unbiased output. 
The method is simple to implement with reference to \cite[Theorem 2.3]{https://doi-org.utokyo.idm.oclc.org/10.1002/rsa.20654}, only requiring the generation of iid standard exponential random variables in parallel with the iid realizations $\{\mathbbm{1}(g(X_k)\le 0)\}_{k\in\mathbb{N}}$. 

Yet another relative-precision stopping rule, introduced in \cite{LWP2013}, is designed to operate under more general conditions than iid Bernoulli trials: (i) $X_n$ is $\mathcal{F}_n$-measurable for a suitable filtration $(\mathcal{F}_n)_{n\in\mathbb{N}_0}$, (ii) $\mathbb{E}[X_n|\mathcal{F}_{n-1}]=\theta$ for some $\theta>0$, and (iii) $X_n\in [0,1]$ with probability one, for all $n\in \mathbb{N}$. 
In the present context, with the simulation length defined by $n_r:=\{n\in \mathbb{N}:\, nP_f(n)\ge r\}$, various properties of the error probability $\mathbb{P}(
P_f(n_r)-P_f
/P_f>\epsilon)$ and the expected simulation length $\mathbb{E}[n_r]$ are derived, allowing for the (albeit nontrivial) selection of the threshold $r$ to achieve a desired pair $(\epsilon,\delta)$. 

\textcolor{black}{Despite these advances, a fundamental question arises as to whether strict $(\epsilon,\delta)$-guarantees can be achieved at all in the absence of prior knowledge of $P_f$.
The reliance on auxiliary estimation steps suggests that fully non-asymptotic guarantees may be unattainable without embedding additional assumptions or accepting conservative bounds.
In particular, existing workaround strategies highlight a deeper tension.
Either one assumes partial prior information (for instance, lower bounds on $P_f$) to maintain formal guarantees, or one adopts adaptive procedures whose guarantees become implicit or probabilistically conditioned on preliminary estimates.
This observation points to a fundamental trade-off between theoretical rigor and practical implementability, raising the open question of whether strict probabilistic guarantees in the $(\epsilon,\delta)$ framework are achievable under realistic conditions without prior knowledge of $P_f$.}

Error-based stopping criteria above, mainly developed outside the immediate domain of structural reliability, provide a rigorous probabilistic foundation that is often missing in structural reliability analysis. By framing stopping rules in terms of relative error tolerances and confidence guarantees, such as the $(\epsilon,\delta)$-approximation of the failure probability, these approaches ensure explicit control over both estimation accuracy and reliability of inference. 
\textcolor{black}{At present, however, the extent to which such guarantees can be maintained without incurring prohibitive computational costs in the small-probability regime remains an open theoretical issue, pointing to a fundamental trade-off between statistical rigor and computational efficiency.} 
Integrating these principles into structural reliability analysis would not only elevate the methodological rigor of stopping criteria but also have the potential to bridge the gap between practical engineering applications and the deeper probabilistic guarantees available in the broader literature.

\textcolor{black}{In summary, despite its strong theoretical guarantees, this framework is rarely used in its strict form in structural reliability practice, primarily due to its dependence on unknown quantities and the high computational cost in rare-event estimation.
}

\subsection{Absolute error-based stopping criteria}
\label{subsection absolute error MC}

In contrast to the relative error-based criterion \eqref{relative error}, consider the stopping criterion based on the absolute error, as follows:
\begin{equation}\label{primitive definition of stopping rule}
 \tau_{\epsilon,\delta}\leftarrow \argmin_{n\in \mathbb{N}}\left\{\mathbb{P}\left(|P_f(n)-P_f|>\epsilon\right)\le \delta\right\}.
\end{equation}
In words, the simulation terminates once the probability of the estimate $P_f(n)$ based on $n$ iterations is far from the (unknown) probability of failure $P_f$ by the error tolerance $\epsilon$ or more is less than or equal to the significance level $\delta$.

In the literature on structural reliability analysis, absolute error has rarely been employed as a basis for stopping criteria.
A likely reason is the challenge of specifying an appropriate tolerance $\epsilon$, since the magnitude of the failure probability $P_f$ is typically unknown in advance.
Nonetheless, to encourage further developments in this area, we examine a range of absolute error-based stopping rules, many of which trace their origins to other fields.

\subsubsection{Chebyshev's inequality}
\label{subsubsection chebyshev}

As done in Section \ref{subsection error-based stopping criteria}, apply the Chebyshev inequality to the error probability \eqref{primitive definition of stopping rule}, as follows:
\[
\mathbb{P}(|P_f(n)-P_f|> \epsilon)\le \frac{P_f(1-P_f)}{n\epsilon^2},\quad n\in \mathbb{N},
\]
which yields the desired criterion $\mathbb{P}(|P_f(n)-P_f|> \epsilon)\le \delta$ for all $n\ge \lceil P_f(1-P_f)/(\delta \epsilon^2)\rceil=:n_{\rm Cheb0}$.
This is not entirely practical unless the value of $P_f(1-P_f)$ is available in advance.
With the aid of the upper bound $P_f(1-P_f)\le 1/4$ for all $P_f\in [0,1]$, we may instead achieve the desired criterion $\mathbb{P}(|P_f(n)-P_f|>\epsilon)\le \delta$ for all $n\ge n_{\rm Cheb1}:=\lceil 1/(4\delta \epsilon^2)\rceil\ge \lceil P_f(1-P_f)/(\delta \epsilon^2)\rceil=n_{\rm Cheb0}$.

The threshold $n_{\rm Cheb1}$ is clearly more conservative than $n_{\rm Cheb0}$, as $P_f(1-P_f)\approx P_f\ll 1/4$ for a small failure probability (such as $P_f=10^{-10}$).
If an upper bound for $P_f$ is known, this rough bound of $1/4$ can be refined.
For example, if $P_f \le c$ for some $c \in (0,1/2)$, then $P_f(1-P_f) \le c(1-c) < 1/4$, leading to the threshold $\lceil c(1-c)/(\delta \epsilon^2)\rceil$.
This refinement improves $n_{\rm Cheb1}$, though it remains less optimal than $n_{\rm Cheb0}$.

We note that, as rigorously established in \cite{RL1956}, any stopping rule that satisfies the criterion \eqref{primitive definition of stopping rule} requires prior information beyond the sampled data, thereby necessitating a framework of two-stage procedures.
The specific framework considered here has a unique structure of iid Bernoulli trials with a known upper bound on the theoretical variance, that is, $P_f(1-P_f)\le 1/4$.
This eliminates the need for a two-stage procedure.

\subsubsection{Central limit theorem}
\label{subsubsection CLT}

Instead of ensuring the desired stopping criterion \eqref{primitive definition of stopping rule}, typical stopping rules have long been based on the central limit theorem \eqref{CLT1}, despite lacking such a guarantee, such as 
\begin{equation}\label{primitive CLT approximation}
\mathbb{P}\left(\left|P_f(n)-P_f\right|>\epsilon\right)=\mathbb{P}\left(\sqrt{n}\frac{|P_f(n)-P_f|}{\sigma(n)}>\sqrt{n}\frac{\epsilon}{\sigma(n)}\right)\sim 2\left(1-\Phi\left(\sqrt{n}\frac{\epsilon}{\sigma(n)}\right)\right)\le \delta.
\end{equation}
In light of the expected convergence $\sigma^2(n)\to P_f(1-P_f)$ and the upper bound $P_f(1-P_f)\le 1/4$, one obtains the threshold $n_{\rm CLT}:=\lceil z_{\delta/2}/(4\epsilon^2)\rceil$ based on the rough estimate $\sigma^2(n)\approx 1/4$, resulting in $\mathbb{P}(|P_f(n)-P_f|>\epsilon)\lessapprox \delta$ for $n\ge n_{\rm CLT}$.
Here, we have denoted by $z_{\delta/2}$ the $100(1-\delta/2)\%$-quantile of the standard normal distribution (that is, $1-\delta/2=\mathbb{P}(Z\le z_{\delta/2})$ for $Z\sim \mathcal{N}(0,1)$).
Let us stress that, unlike those in Section \ref{subsubsection chebyshev}, the threshold $n_{\rm CLT}$ here does not guarantee the desired criterion $\mathbb{P}(|P_f(n)-P_f|>\epsilon)\le \delta$, since the stopped empirical variance $\sigma^2(n_{\rm CLT})$ can be much smaller than the applied upper bound $1/4$.

Instead of approximating $\sigma^2(n)\approx 1/4$ outright, the stopping criterion can be reformulated following the pioneering work \cite{YH1965}, based on the last two terms in the approximation \eqref{primitive CLT approximation} retaining the empirical variance $\sigma^2(n)$, as follows: 
\begin{equation}\label{def of tau}
\tau_{\epsilon,\delta}:=\argmin_n\left\{z_{\delta/2}\frac{\sigma(n)}{\sqrt{n}} \le \epsilon\right\},
\end{equation}
where the arguments of the minima can be taken over all natural numbers (that is, $\argmin_{n\in \mathbb{N}}$), or restricted to natural numbers exceeding a specified threshold (that is, $\argmin_{n\in \{n_{\min},n_{\min}+1,\cdots\}}$ for some predetermined $n_{\min}\in \mathbb{N}$), depending on the nature of the problem at hand.

As a fundamental requirement for stopping criteria, small values of the error tolerance $\epsilon$ need to correspond to large values of simulation length $\tau_{\epsilon,\delta}$, that is, $\tau_{\epsilon,\delta}\to +\infty$ with probability one as $\epsilon\to 0$.
In light of the formulation \eqref{def of tau}, this requirement is clearly equivalent to the convergence of the normalized standard deviation, that is, 
$\sigma(n)/\sqrt{n}\to 0$ with probability one as $n\to +\infty$.
Next, the formulation \eqref{def of tau} yields the inequalities $\epsilon\sqrt{\tau_{\epsilon,\delta}-1}/\sigma(\tau_{\epsilon,\delta}-1)<z_{\delta/2}\le \epsilon\sqrt{\tau_{\epsilon,\delta}}/\sigma(\tau_{\epsilon,\delta})$, resulting in the convergence $\epsilon\sqrt{\tau_{\epsilon,\delta}}/\sigma(\tau_{\epsilon,\delta})\to z_{\delta/2}$ as $\epsilon\to 0$,
provided $\tau_{\epsilon,\delta}\to  +\infty$ and $\sigma(n)/\sqrt{n}\to 0$, as discussed.

A key issue of the formulation \eqref{def of tau}, particularly in practice, is that an excessive number of zeros in the empirical variance can persist for an exceptionally long period.
To delve into this problem a little further, the experimenter observes $P_f(n)=0$ for $n\in \{1,2,\cdots,N_1-1\}$ during some very long time $N_1$, which follows the geometric distribution $\mathbb{P}(N_1=k)=(1-P_f)^{k-1}P_f$ for $k\in \mathbb{N}$, with $\mathbb{E}[N_1]=1/P_f$ and ${\rm Var}(N_1)=(1-P_f)/P_f^2$.
In this situation, we have $(P_f(1),P_f(2),\cdots,P_f(N_1-1),P_f(N_1))=(0,0,\cdots,0,1/{N_1})$, followed by a probabilistic restart.
That is, one needs to wait for another geometric length $N_2$ until the second success.
In general, with $m$ representing the number of successes and the notation $M_m:=N_1+\cdots +N_m$, we have $P_f(M_0+n)=0$ for $n\in \{1,\cdots,N_1-1\}$ and $P_f(M_m+n)=m/(M_m+n)$ for $(m,n)\in \mathbb{N}\times \{0,1,\cdots,N_{m+1}-1\}$.
Accordingly, the empirical variance remains zero until the first success as well, resulting in 
    \begin{equation}\label{too many zeros in s2}
\left(\sigma^2(2),\sigma^2(3),\cdots,\sigma^2(N_1-1),\sigma^2(N_1)\right)=\left(0,0,\cdots,0,\frac{N_1-1}{N_1^2}\right),
    \end{equation}
    and then, generally, $\sigma^2(M_m+n)=m(M_m+n-m)/((M_m+n-1)(M_m+n))$ 
    for $(m,n)\in \mathbb{N}\times \{0,1,\cdots,N_{m+1}-1\}$.
By the strong law of large numbers $M_m/m\to \mathbb{E}[N_1]=1/P_f$ as long as $P_f>0$, it holds that
\begin{equation}\label{bernoulli convergence}
 P_f(M_m+n)=\frac{m}{M_m+n}\to P_f, \quad 
 \sigma^2(M_m+n)=\frac{m(M_m+n-m)}{(M_m+n-1)(M_m+n)}\to P_f(1-P_f),
\end{equation}
with probability one as $m\to +\infty$.
Let us stress that the convergences \eqref{bernoulli convergence} hold true, no matter how small the theoretical mean $P_f$ is (like even when it is on the order of $\mathcal{O}(10^{-10})$).

The stopping criterion \eqref{def of tau} often results in the fixed minimum length $n_{\min}$ (that is, $\tau_{\epsilon,\delta}=n_{\min}$), leading to $P_f(\tau_{\epsilon,\delta}) = P_f(n_{\min}) = 0$ and $\sigma^2(\tau_{\epsilon,\delta}) = \sigma^2(n_{\min}) = 0$ when $N_1 > n_{\min}$.
Specifically, as indicated in the successive empirical variances \eqref{too many zeros in s2}, the empirical variance remains zero until the first geometric time $N_1$, which, on average, is $\mathbb{E}[N_1] = 1/P_f$.
To prevent premature termination, one could set $n_{\min}$ to $\mathcal{O}(1/P_f)$, though $P_f$ is unknown. 
Alternatively, a deterministic inflation of the interval width \cite{PW1992} using a monotonically decreasing sequence ${a_n}{n\in\mathbb{N}}$ can be employed, based on the inequality $\argmin_n\{z_{\delta/2}\sigma(n)/\sqrt{n}+a_n\le \epsilon\}\ge \argmin_n\{a_n\le \epsilon\}\to +\infty$ as $\epsilon\to 0$.
However, similar to the fixed minimum length, the sequence ${a_n}{n\in\mathbb{N}}$ must be carefully chosen to ensure that ${a{N_1} > \epsilon}$ holds with high probability. Otherwise, the stopping rule is likely to produce $P_f(\tau_{\epsilon,\delta}) = 0$.

We note that, unlike counting the number of successes, a stopping rule is examined in \cite{LJ2006} where the process continues until a predetermined number of successes is reached.
This approach results in a negative binomial distribution for the total number of trials.
Regardless of whether the stopping condition is based on a random number of successes or trials, if the experimenter retains the ability to retrospectively resume the procedure, the framework falls under optional stopping \cite{fletcher2024, OHKUBO2021}, which is not the intended setting here.

\textcolor{black}{To sum up, these criteria are sometimes adopted in reliability estimation when asymptotic approximations are acceptable, while their validity is often questionable in structural reliability analysis, particularly for small sample sizes or highly skewed and rare-event distributions.}

\subsubsection{Hoeffding's inequality}
\label{subsubsection hoeffding}

Instead of Chebyshev's inequality (Section \ref{subsubsection chebyshev}) or the central limit theorem (Section \ref{subsubsection CLT}), one can apply Hoeffding's inequality $\mathbb{P}(|P_f(n)-P_f|\ge \epsilon)\le 2e^{-2n\epsilon^2}$ to guarantee the criterion $\mathbb{P}(|P_f(n)-P_f|> \epsilon)\le \delta$ for all $n\ge \lceil \ln(2/\delta)/(2\epsilon^2)\rceil=:n_{\rm Hoeff}$.
Compared to the previous threshold $n_{\rm CLT}$ (derived in Section \ref{subsubsection CLT}), the threshold $n_{\rm Hoeff}$ provides an improvement with their ratio given by $n_{\rm Hoeff}/n_{\rm CLT}\approx 2\ln (2/\delta)/\Phi^{-1}(1-\delta/2)$, which exceeds $3.5$ for $P_f\le 0.1$.
Nonetheless, even this refined threshold has been reported to remain conservative for extremely small failure probability $P_f$ \cite[Figure 1]{Jiang2014}.

\textcolor{black}{Let us note that although these bounds provide non-asymptotic guarantees, they tend to be overly conservative in practice and are therefore infrequently applied in structural reliability analysis.}

\subsection{Stability-based stopping criteria}
\label{subsection stability-based stopping criteria}

A stopping criterion is developed in \cite{WOO1991179} based upon the stability of the probability estimates, measured through their successive change $|P_f(n+1)-P_f(n)|$, in the form of
\begin{equation}\label{successive changes}
 \tau_{\epsilon,\delta}\leftarrow \argmin_{n\in\mathbb{N}}\left\{\mathbb{P}(|P_f(n+1)-P_f(n)|>\epsilon)<\delta\right\},
\end{equation}
which is conceptually different from the one specified in the conventional framework \eqref{primitive definition of stopping rule}.
In the stopping rule \eqref{successive changes}, the error tolerance $\epsilon$ represents the maximum permitted change in the estimate $P_f(n)$ resulting from an additional sample, as opposed to the conventional framework \eqref{primitive definition of stopping rule}, where the error tolerance $\epsilon$ denotes the maximum permitted distance of the estimate $P_f(n)$ from the unknown theoretical mean $\mu$.

The simulation length $\tau_{\epsilon,\delta}$ can then be derived \cite{Saw01051984} based on the criterion \eqref{successive changes} with the aid of an inequality of Chebyshev type with respect to the $(n+1)$st realization based on the empirical standard deviation $\sigma(n)$, as given in the form of 
\[
\mathbb{P}\left(\left|\mathbbm{1}(g(X_{n+1})\le 0)-P_f(n)\right|>\lambda \sqrt{(n+1)/n}\sigma(n)\right)\le \frac{\lfloor (n+1)(n-1+\lambda^2)/(n\lambda^2)\rfloor}{n+1},
\]
for all $\lambda \ge 1$ and $n\in \mathbb{N}$.
We note that the stopping criterion with reference to the successive change, whether absolute or relative, has often appeared in the context of structural reliability analysis \cite{PENG2022103353, ROY2020106948, WANG2022108214}, among others.

Clearly, these criteria presume that stability guarantees convergence to the true value, which is however not necessarily valid (for instance, in the presence of local optima).
\color{black}More specifically, this assumption may fail in structural reliability analysis where the limit state function exhibits multimodality or highly nonlinear behavior.
In such cases, the estimator $P_f(n)$ may appear locally stable while still being biased due to incomplete exploration of the failure domain, particularly when multiple disconnected failure regions exist.

Typical failure modes include (i) premature stabilization around suboptimal regions when sampling fails to discover critical failure modes, (ii) apparent convergence caused by surrogate-induced misclassification of rare failure events, and (iii) oscillatory or plateau behavior in which small successive changes mask slow global drift toward the true value. These effects are especially pronounced in high-dimensional or rare-event settings, where the probability mass is concentrated in narrow and difficult-to-detect regions.

In summary, these observations suggest that stability should not be interpreted as a sufficient indicator of correctness, but rather as a local convergence signal whose reliability depends on the global structure of the failure domain and the adequacy of sampling or surrogate exploration.
In practice, despite stability-based criteria are widely used due to their simplicity, they are typically employed alongside other measures to mitigate their sensitivity to local convergence artifacts.

\subsection{Synthesis and theoretical perspective}
\label{subsection summary 3}

Those classical stopping criteria provide a rigorous probabilistic foundation for Monte Carlo simulation, whereas their direct applicability to structural reliability analysis varies.
Variance- and error-based rules offer strong theoretical guarantees but often depend on unknown quantities, which makes their strict implementation challenging in practice.
In rare-event settings, they may require prohibitively large sample sizes, thereby limiting their direct applicability.
Although non-asymptotic guarantees provide alternative assurances, they typically come at the cost of conservativeness or computational inefficiency.
As a consequence, the standard practice is to rely on simplified or heuristic variants, trading strict guarantees for computational feasibility.
These limitations have motivated the development of hybrid and model-based stopping strategies, which are discussed in the following sections.

\color{black}
\section{Stopping criteria for sampling methods}
\label{section sampling methods}

In Section \ref{section stopping criteria for monte carlo simulation}, we have reviewed stopping criteria for Monte Carlo simulation, drawing from both structural reliability analysis and the broader literature, while also highlighting common challenges in developing effective stopping rules for estimating small failure probabilities.
In this section, we focus on stopping criteria for related sampling methods, that is, importance sampling,
subset simulation,
line sampling,
directional sampling,
and Markov chain Monte Carlo,
as presented in Sections \ref{subsection importance sampling}-\ref{subsection MCMC},
each designed to enhance the implementation and accuracy of Monte Carlo-based structural reliability analysis.

\subsection{Importance sampling}
\label{subsection importance sampling}

Importance sampling (IS) is a variance-reduction technique widely utilized in structural reliability to estimate small failure probabilities efficiently \cite{TABANDEH2022102216}.
It involves sampling from an alternative probability density, often termed the IS density, which places greater emphasis on critical regions of the input space, especially the failure domain.
More precisely, the IS estimate of failure probability takes the form 
\[
P_f(n) = \frac{1}{n}\sum_{k=1}^{n}\mathbbm{1}(g(X_k)<0)\frac{f(X_k)}{h(X_k)},
\]
where the sample points $\{X_k\}_{k\in \mathbb{N}}$ are drawn independently from the IS density $h({\bf x})$, rather than from the original input distribution $f({\bf x})$.
The optimal IS density $h^{\star}({\bf x})$, minimizing estimator variance, is analytically given by $h^\star(\mathbf{x})=\mathbbm{1}(g(\mathbf{x})<0)f(\mathbf{x})/P_f$ in theory, yet practical application of this density is nontrivial because it assumes advance knowledge of the unknown failure probability $P_f$ to be estimated.
Let us emphasize that considerable care is required when selecting the IS density, as an inappropriate choice can readily lead to infinite variances \cite{10.1214/25-PS32}, in which case the second moment diverges (that is, $\int_{\Omega}(\mathbbm{1}(g({\bf x})<0)f({\bf x})/h({\bf x}))^2h({\bf x})d{\bf x}=+\infty$).

Various adaptive methods have thus been developed to iteratively approximate the optimal IS density.
Sequential importance sampling (SIS), proposed in \cite{PAPAIOANNOU201666}, employs a smoothing parameter  to approximate the indicator function, leading to a sequence of gradually refined IS densities.
The key updating criterion involves minimizing the difference between the CoV of weights and a prescribed target value $\sigma_j = \argmin_{\sigma' \in (0,\sigma_{j-1})}\| \widehat{\delta}_{w_j}(\sigma') - \delta_\mathrm{target}\|$.
The iterative process halts once the CoV of the weights falls below the predefined threshold, ensuring sufficient convergence of the IS density.
Sequential directional importance sampling (SDIS) \cite{CHENG2023102291, CHENG2025102574} further refines this approach by 
employing directional sampling.
Their adaptive parameter is similarly updated, and the iterative procedure terminates once the so-called magnification factor reaches unity, ensuring convergence to the true limit state surface.

Cross entropy (CE) IS methods have also been extensively studied in, for instance, \cite{papaioannouI2019, wangZ2016}.
These methods adaptively approximate the optimal IS density through parametric auxiliary densities (often Gaussian mixtures \cite{kurtzN2013}), updated iteratively by minimizing the Kullback-Leibler divergence between the current density and the optimal IS density.
The stopping criterion for CE methods typically involves monitoring quantiles of the limit state function or achieving stability in the estimated failure probability across iterations \cite{geyerS2019, uribeF2021}.
More recently, in the improved cross-entropy importance sampling with non-parametric Gaussian mixture (iCE-NGM) \cite{ZHANG2025111322}, a budget-informed stopping criterion is developed.
Instead of relying on a fixed threshold for the coefficient of variation on the importance weights, the developed stopping rule adaptively balances proposal refinement against the remaining computational budget, by monitoring the budget-constrained coefficient of variation of the final IS estimator.

Incorporation of surrogate models and active learning strategies into IS has further enhanced stopping criteria, such as in \cite{CADINI2020101998, DUBOURG201347, JIA2024102427}.
Predictive error estimates or expected improvements are adopted in \cite{YUN2020101891, YUN2021103131} to establish robust stopping conditions, optimizing computational efficiency by focusing sampling where uncertainty or expected gain is highest.
Variance-based sensitivity analyses offer quantitative stopping criteria through total coefficients of variation (TCOV), providing statistical assurance of estimator stability \cite{MENZ2021102116}.
Bayesian updating strategies in IS, explored in \cite{XIAO2022102230}, employs normalized effective sample size to measure the similarity between the IS density and the target density.
Moreover, advanced IS variants, such as ensemble Kalman filters \cite{wagnerF2022}, Hamiltonian MCMC \cite{PAPAKONSTANTINOU2023103485}, 
Langevin importance sampling \cite{TABANDEH2026111634}, consensus-based approaches \cite{doi:10.1137/23M1565966}, and seminal adaptive schemes \cite{auSK1999, BUCHER1988119} propose novel convergence diagnostics, significantly broadening the methodological landscape.

\subsection{Subset simulation}
\label{subsection subset simulation}

The subset simulation method, originally introduced by Au and Beck \cite{AU2001263}, is widely used in structural reliability analysis to efficiently estimate small failure probabilities. The central principle behind subset simulation involves expressing a rare failure event as a sequence of more probable conditional events.
Specifically, let $F:=\{g(X)\leq 0\}$ be the event of failure and denote by $\{F_k\}_{k\in \{1,\cdots,m\}}$ a sequence of sub-events, satisfying the nesting condition $F_1\supset F_2\supset \cdots \supset F_m=F$ and $F_k=\cap_{j=1}^kF_j$ for all $k\in\{1,\cdots,m\}$.
Then, the probability of the rare event $F$ can be decomposed into a product of conditional probabilities by conditioning on the sub-events $\{F_k\}_{k\in \{1,\cdots,m-1\}}$ sequentially:
\[
	P_f=\mathbb{P}(F_m)=\mathbb{P}(F_1)\frac{\mathbb{P}(F_2)}{\mathbb{P}(F_1)}\frac{\mathbb{P}(F_3)}{\mathbb{P}(F_2)}\cdots \frac{\mathbb{P}(F_m)}{\mathbb{P}(F_{m-1})}=\mathbb{P}(F_1)\mathbb{P}(F_2|F_1)\mathbb{P}(F_3|F_2)\cdots\mathbb{P}(F_m|F_{m-1}),
\]
This approach effectively transforms a challenging rare-event simulation into a sequence of simulations involving more frequent intermediate events, significantly improving computational tractability.

In practice, the procedure starts with a standard Monte Carlo estimation of $\mathbb{P}(F_1)$, the probability of the initial intermediate event. Subsequently, conditional samples are often simulated using Markov chain Monte Carlo (Section \ref{subsection MCMC}), where the stationary distribution of the generated Markov chain at each subset level corresponds to the conditional density:
\[
f_k(\mathbf{x}) = \frac{f(\mathbf{x})\mathbbm{1}(g(\mathbf{x})\leq c_k)}{\mathbb{P}(F_k)}, \quad F_k:=\mathbbm{1}\left(g(X)\le c_k\right).
\]
The intermediate thresholds $c_1 > c_2 > \cdots$ are selected recursively to ensure that each conditional probability remains close to a fixed value, typically chosen as $0.1$ in practice \cite{Zuev2021}.

A critical aspect of subset simulation involves determining an appropriate stopping criterion. Standard practice involves monitoring the conditional probability estimates $\mathbb{P}(F_k|F_{k-1})$ at each subset level. The iterative procedure continues until the conditional probability at level $k$ surpasses the predetermined threshold $p$. At this stage, the final level is established, and the failure probability is approximated as $P_f \approx p^{m-1}\mathbb{P}(F_m|F_{m-1})$, using the estimated conditional probability from the final iteration \cite[Section C]{Zuev2021}.
Alternatively, the process advances to the next intermediate subset if the conditional probability has not yet exceeded the threshold.

Beyond this basic criterion, various advanced stopping rules have been explored to enhance subset simulation efficiency and accuracy.
Recent developments often incorporate computationally adaptive techniques, such as mesh refinement \cite{doi:10.1137/22M1515240}, and sophisticated partitioning strategies for the input space \cite{KINNEAR2025103729}. Such methods adaptively control the granularity of simulations based on statistical convergence metrics, variance stabilization, or spatial accuracy in surrogate modeling.
We direct the reader to \cite{GIOVANIS2022103293} for a detailed treatment of uncertainty quantification in failure probability estimates arising from incomplete knowledge of input distributions, and to \cite{SHIELDS2021106431} for cases involving strongly non-Gaussian, highly anisotropic, and degenerate distributions, where the estimation of the CoV has been refined by accounting for the cross-correlation effects among multiple Markov chains at each conditional level, thereby mitigating the underestimation issue \cite{AU2007183}.

Significant progress has also been made by integrating subset simulation with surrogate models and active learning strategies, particularly Gaussian process surrogate models and Bayesian active learning \cite{doi:10.1137/16M1078276, DUBOURG2011, HUANG201686, WALTER201510, ZHANG2022103256}.
These approaches adopt posterior uncertainty quantification and predictive variance metrics to formulate stopping criteria.
Machine learning methods, such as ANN-based surrogate models \cite{CUI201912}, have been combined with subset simulation to enhance sampling efficiency and terminate simulations based on surrogate prediction accuracy.
Subset simulation is reformulated within a multifidelity framework in \cite{NADERI2026111739} by modeling the fidelity level as a discrete random variable, which allows adaptive sampling across different fidelities and achieves a more efficient allocation of computational effort between low- and high-fidelity models.
Moreover, although the proposed partial ring simulation \cite{LENG2026102648} is not a direct variant of subset simulation, it embodies a closely related concept of sequentially exploring the failure domain using MCMC sampling and probability-based stopping criteria.
These advancements and extensions have considerably expanded the scope of subset simulation, yet calling for stopping criteria that are systematic and computationally efficient.

\subsection{Line sampling}
\label{subsection line sampling}

Line sampling aims to estimate the failure probability by sampling along lines, rather than individual random points.
In particular, it samples on a hyperplane perpendicular to a dominant direction pointing toward the failure domain.
This approach remains effective even in high-dimensional settings, provided that such dominant directions can be reliably identified \cite{Pradlwarter_2007, Schueller_Pradlwarter_Koutsourelakis_2004, ZIO20101300}.

To formulate the method in brief, consider a unit vector $\boldsymbol{\alpha}$ that identifies an important direction on the unit hypersphere $S^{d-1}$, typically pointing toward the nearest part of the failure region to the origin.
Let the failure domain in the standard Gaussian space be defined by $F:=\{{\bf z} \in\mathbb{R}^d:g_T({\bf z})\leq 0\}$.
Suppose the standard normal vector ${\bf z}:=(z_1,z_2,\cdots,z_d)$ is oriented such that its first component $z_1$ aligns with the direction of the unit vector $\boldsymbol{\alpha}$.
In addition, assume that a mapping $G_0:\mathbb{R}^{d-1}\to \mathcal{P}(\mathbb{R})$ is well defined, by which we denote the power set of $\mathbb{R}$, that allows the failure set to be expressed as $F=\{{\bf z}\in\mathbb{R}^d: z_1\in G_0(z_2,\cdots,z_d)\}$.
Assuming a single-entry point per line into the failure domain, the failure probability in the integral form \eqref{reliability gaussian formulation} can be reformulated in the following way:
\begin{align}
P_f	=\int_F\prod_{k=1}^d \phi(z_k)dz_k
	&=\int_{\mathbb{R}^{d-1}}\left(\int_{G_0(z_2,\cdots,z_d)}\phi(z_1)dz_1\right)\prod_{k=2}^d \phi(z_k)dz_k\nonumber\\
&=\int_{\mathbb{R}^{d-1}}\Phi(G_0(z_2,\cdots,z_d))\prod_{k=2}^d \phi(z_k)dz_k=\mathbb{E}[\Phi(G_0(Z))],\label{LS formulation}
\end{align}
with $\Phi(D):=\int_{D}\phi(z)dz$ for $D\subseteq\mathbb{R}$ and $Z$ denoting the $(d-1)$-dimensional standard Gaussian random vector in the rightmost expectation.
MCS based on the last integral representation in the formulation \eqref{LS formulation} may reduce the estimator variance \cite{Koutsourelakis_Pradlwarter_Schuller_2004}.
The reduction in estimator variance observed in this context stems from lowering the dimensionality of the underlying random variable by projecting the $d$-dimensional standard Gaussian distribution onto a $(d-1)$-dimensional subspace, as established in the formulation \eqref{LS formulation}, unless the function $G(z_1,\cdots,z_d)$ is independent of the first component $z_1$.
For further discussion and theoretical justification of line sampling, particularly regarding the convergence behavior of the resulting estimators, we direct the reader to \cite{Koutsourelakis_2004, Zio_Pedroni_2010}.
It is worth noting that the effectiveness of line sampling deteriorates sharply when multiple disconnected failure regions arise, reflecting that the single-entry assumption is intrinsic to the above formulation and validity range (for instance, \cite{Pradlwarter_2007}).

The implementation of the standard line sampling method involves two main steps: selecting the important direction, followed by applying MCS to the last integral representation in the formulation \eqref{LS formulation}.
A few studies have proposed adaptive strategies for identifying the important direction.
For instance, in \cite{PAPAIOANNOU2021102025}, the direction is iteratively updated as the distance from the origin to the failure surface along this direction decreases.
For the second step, the CoV is commonly adopted as the primary convergence criterion (see, for instance, \cite{Koutsourelakis_2004, Koutsourelakis_Pradlwarter_Schuller_2004, Shayanfar_2017_2}), just as in the standard MCS (Section \ref{subsection CoV}).
It has often been reported that line sampling leads to a remarkably stable CoV across a wide spectrum of reliability problems, largely independent of the actual failure probability \cite{Zio_Pedroni_2009}.
We note that recent advancements in stopping criteria for line sampling have frequently appeared alongside other frameworks, such as active learning reliability methods (Section \ref{subsection active learning}) and Bayesian methods (Section \ref{subsection bayesian methods}), as investigated in  \cite{DANG2023102351, DANG2023116068, DANG2024110053, SONG2021107113}, to name just a few.

\subsection{Directional sampling}
\label{subsection directional sampling}

Directional sampling is a variance reduction technique in structural reliability analysis that estimates the failure probability by sampling over directions on the unit hypersphere.
In polar coordinates, the random vector $Z$ can be decomposed as $Z = R V$, where $V = Z/\|Z\|$ lies on the unit hypersphere $S^{d-1} = \{{\bf v} \in \mathbb{R}^d : \|{\bf v}\| = 1\}$, and $R = \|Z\|$ is the radial distance in $(0, +\infty)$.
When the random vector $Z$ follows the standard multivariate normal distribution, $V$ is uniformly distributed on $S^{d-1}$, and $R^2$ follows a chi-square distribution with $d$ degrees of freedom, where $V$ and $R$ are independent.
The failure probability can then be expressed as an integral over the unit hypersphere \cite{Bjerager_1988}:
\[
	P_f=\int_{S^{d-1}}\mathbb{P}\left(G(RV)\leq 0 | V={\bf v}\right)f_{V}({\bf v})d{\bf v}
	=\int_{S^{d-1}}\mathbb{P}\left(G(R{\bf v})\leq 0 \right)f_{V}({\bf v})d{\bf v},
\]
where $f_V({\bf v}) = \Gamma(d/2)/(2\pi^{d/2})$ denotes the uniform density on $S^{d-1}$.
Under the global monotonicity condition, the failure probability along a given direction ${\bf v}$ reduces to the tail probability of $R$ exceeding a threshold $r({\bf v})$, namely, $\mathbb{P}(G(R{\bf v}) \leq 0) = \mathbb{P}(R \geq r({\bf v})) = 1 - F_{\chi^2_d}(r^2({\bf v}))$, where $F_{\chi^2_d}$ denotes the cumulative distribution function of the chi-square distribution with $d$ degrees of freedom.
In such cases, the directional sampling estimator can lead to much smaller variance than relying on crude Monte Carlo simulation by deterministically computing $r(V)$ along the sampled direction $V$ and suppressing randomness in the random variable $R$.

Designing effective stopping criteria for directional sampling is crucial for balancing computational cost and estimation accuracy.
A widely adopted strategy is to monitor the CoV (Section \ref{subsection CoV}) of the estimator.
The sampling process is terminated once the CoV drops below a predefined value, indicating sufficient statistical precision \cite{GROOTEMAN2011134, SHAYANFAR201814, ASDIS, ReliabilitySS}.

Despite its notable advantages in variance reduction, directional sampling often depends on the assumption of global monotonicity of the limit state function along sampled directions.
When this assumption is violated, as in highly nonlinear or nonmonotonic systems, the method must rely on approximate root-finding procedures or surrogate-based estimations of the threshold distance in each direction.
Under such conditions, stopping criteria must be carefully formulated to guarantee not only statistical convergence but also to incorporate the effects of surrogate approximation errors, numerical instability in root-finding, and directional uncertainty propagation.
It appears that the development of robust stopping rules that can relax the monotonicity assumptions remains largely unexplored, representing a significant and timely research frontier in directional sampling.

\subsection{Markov chain Monte Carlo}
\label{subsection MCMC}

In structural reliability analysis, Markov chain Monte Carlo (MCMC) is a widely used sampling method for estimating failure probabilities by simulating the behavior of uncertain parameters in complex structural systems.
It is employed not only within Bayesian frameworks (Section \ref{subsection bayesian methods}) but also as a key component of subset simulation (Section \ref{subsection subset simulation}).

Stopping criteria for the MCMC component are, on the one hand, not always explicitly formulated in structural reliability analysis.
Many contributions instead emphasize efficient samplers or variance reduction strategies under fixed simulation budgets.
For instance, in \cite{PAPAIOANNOU201589}, subset simulation is enhanced with conditional and adaptive MCMC schemes to reduce correlation and improve acceptance rates, yet no stopping rules are prescribed beyond fixed sample sizes.
Similarly, in the sequential and adaptive probabilistic integration (SAPI) method \cite{KITAHARA2025102577}, the MCMC subroutine is run with fixed chain lengths and burn-in assumed negligible, so no explicit adaptive stopping criterion is defined for the chains themselves.

On the other hand, several works incorporate explicit termination rules.
A two-stage procedure \cite{XIAO2022107938} stops its first stage once the chain reaches the failure domain, and its second after a predefined number of accepted samples, with adaptive tuning of acceptance rates.
Likewise, in \cite{SONG2023711}, the MCMC exploration is terminated in the first stage once the majority of samples are classified with more than 95\% probability of being in the failure domain.

In contrast to the limited developments of stopping criteria formulated specifically for MCMC components in structural reliability analysis, the field of computational statistics offers a mature and comprehensive foundation for MCMC termination.
A vast body of research has established rigorous convergence diagnostics and sequential stopping strategies aimed at ensuring both the accuracy and efficiency of posterior inference (see, for instance, \cite{craiu2025handbook}).
Those established stopping rules has the potential to enhance the stability and computational efficiency, while such integration would require careful attention, since reliability analyses often deal with rare failure events, multimodal response surfaces, and intricate limit state structures, where conventional convergence checks may fail to function properly without appropriate adaptation.

\color{black}
\subsection{Synthesis and methodological insights}

Compared with classical Monte Carlo simulation (Section \ref{section stopping criteria for monte carlo simulation}), sampling-based reliability methods introduce additional layers of complexity in the formulation of stopping criteria.
While many approaches retain variance- or coefficient-of-variation-based rules as core convergence indicators, their interpretation becomes method-dependent, reflecting the specific structure of the sampling strategy.
For instance, in importance sampling and its variants, stopping criteria are closely tied to weight stability and effective sample size, whereas subset simulation relies on intermediate conditional probabilities and level-wise convergence.
In line and directional sampling, convergence is often associated with variance reduction along selected directions, while in Markov chain Monte Carlo, stopping is further complicated by dependence structures and the need for convergence diagnostics.

Across these methods, stopping criteria are rarely universal and are often embedded within the algorithmic design.
As a consequence, they frequently combine multiple indicators, such as variance measures, stability of probability estimates, or budget constraints, rather than relying on a single theoretically grounded rule.
This reflects a broader shift from explicit probabilistic guarantees toward pragmatic convergence assessment tailored to each method.
In particular, issues such as rare-event regimes, high dimensionality, and multimodal failure domains can significantly affect the reliability of standard criteria, requiring additional safeguards or hybrid strategies.

From a practical standpoint, criteria based on coefficient-of-variation remain the most widely used due to their simplicity and interpretability, although their robustness varies across methods.
Stability-based rules are also commonly adopted, especially in iterative or adaptive schemes, yet they should be interpreted cautiously, as apparent convergence may not ensure accurate exploration of the failure domain.
These considerations highlight the need for carefully designed, method-aware stopping strategies, and motivate the transition toward surrogate-based and adaptive approaches discussed in the following sections.

\color{black}
\section{Stopping criteria for advanced modeling frameworks}
\label{section advanced modeling frameworks}

We have reviewed the concept of stopping criteria for Monte Carlo simulation, first in a general context from the perspective of applied probability and statistics (Section \ref{section stopping criteria for monte carlo simulation}), and then in terms of existing criteria  within structural reliability analysis, organized by related sampling methods (Section \ref{section sampling methods}).

We next turn to address stopping criteria for advanced frameworks outside the core realm of Monte Carlo simulation. 
While those stopping criteria can be classified in various ways, one natural approach is to organize them by the types of modeling frameworks.
Along this line, the subsections below discuss stopping criteria associated with Kriging models,
response surfaces,
polynomial chaos expansions,
support vector machines,
neural networks,
multifidelity models,
reduced-order models,
and ensemble models,
as respectively presented in Sections \ref{subsection kriging models}-\ref{subsection ensemble models}.

We note that many references can appear repeatedly across these subsections due to the combined use of relevant models, and reemerge in Section \ref{section advanced methodological frameworks}, where the discussion is structured around specific methodological frameworks to complement the upcoming section.
With this perspective, we hereafter aim to outline the broader landscape of stopping criteria across diverse frameworks, rather than explore the intricate technicalities of individual models, so as to maintain a coherent narrative and highlight overarching trends.
In particular, for in-depth discussions of specific models, we refer the reader to relevant reviews and benchmark studies existing in the literature (Section \ref{subsection existing reviews and benchmark studies}).

\subsection{Kriging models}
\label{subsection kriging models}

Kriging (Gaussian process) surrogates have become a cornerstone of reliability analysis, particularly for mitigating the computational cost of direct Monte Carlo simulation.
Adaptive Kriging-based methods refine the surrogate iteratively by adding new samples in regions of high prediction uncertainty or near the limit state surface.
A key challenge is to define stopping criteria that achieve an effective trade-off between computational efficiency and surrogate accuracy.

To direct the enrichment procedure, criteria built on learning functions are extensively applied, aiming to pinpoint points that most effectively increase surrogate fidelity and the precision of failure probability estimation.
For instance, in \cite{ECHARD2013232, hu_mahadevan_2016, HUANG201686}, the procedure terminates once the $U$-function exceeds 2 for all candidate points, indicating high confidence in surrogate predictions away from the limit state.
Similarly, the expected feasibility function (EFF) \cite{bichon_efficient_2008,WEN2016170} is used as a stopping indicator, with the process ending when the maximum EFF value drops below a specified tolerance.
We defer a more comprehensive treatment of learning functions and their variants to Sections \ref{subsubsection learning functions} and \ref{subsubsection learning Bayesian}.

Several frameworks integrate Kriging-based stopping rules with optimization schemes.
The mixed efficient global optimization (m-EGO) method \cite{hu_du_2015} enriches the surrogate until both the expected improvement (EI) falls below a tolerance and the coefficient of variation (CoV) of the failure probability estimated via AK-MCS \cite{ECHARD2011145} is below a prescribed threshold.
Likewise, the efficient global optimization variants EGO-EI and EGO-AT \cite{Chaudhuri2015} employ EI-based and adaptive targeting criteria, respectively.
In adaptive Kriging-importance sampling (AK-IS) for global sensitivity, stopping combines CoV thresholds with classification confidence \cite{CADINI2020101998}.
Kriging has also been coupled with moving particles \cite{WALTER201510}, where termination is defined by the stabilization of surrogate predictions relative to the failure threshold.
Furthermore, the initial design of experiments can be terminated once a predefined number of failure samples has been identified, guaranteeing that the surrogate sufficiently represents the boundary between safe and failure regions.
Finally, in the AK-MCSi framework \cite{LELIEVRE20181}, a conservative rule is introduced based on the probability that all points in the Monte Carlo population are correctly classified, rather than relying solely on the minimum value of the learning function.

Error-based and stability-based stopping rules constitute an important class.
The error-based stopping criterion (ESC) \cite{wang_ahafieezadeh_2019} derives an explicit bound on the error of the failure probability estimate and terminates enrichment once this bound falls below a prescribed tolerance.
In \cite{yi_zhou_cheng_liu_2020}, stopping is defined by the estimated relative error of the failure probability, with an enhanced bootstrap confidence-based criterion (BSC) further introduced to identify uncertain points prone to status changes and to use their error estimates for robust termination of the active-learning Kriging process.
In the reliability updating with equality information using adaptive Kriging (RUAK) framework \cite{ZHANG2021107323}, stopping criteria are reformulated to control errors in the posterior failure probability.
Unlike conventional error-based rules that only bound the error of the prior failure probability, the proposed two-phase approach first applies the prior error as a preliminary filter, then halts training once the maximum posterior failure probability error is below a user-defined threshold.
Related developments include error-rate-based adaptive Kriging (REAK) \cite{WANG201933}, Bayesian updating with adaptive Kriging (BUAK), which relies on the expected number of misclassified samples among accepted failure points \cite{WANG2020101915}, and probability-adaptive Kriging frameworks that monitor stabilization of the reliability index \cite{KIM2020101924}.
For further variants, we refer the reader to, for instance, \cite{WANG2022108214, YUN2021103131, ZHANG2024109610}.
We revisit error-based and stability-based stopping criteria in Section \ref{subsubsection ALR error}, \ref{subsubsection ALR stability} and Section \ref{subsubsection error stabilization Bayesian}.

Hybrid stopping criteria also arise in ensemble and multi-objective settings.
In ensemble-based learning, budget- and accuracy-oriented rules are combined, using the maximum relative difference between the ensemble-estimated failure probability and that obtained from individual Kriging or ANN models \cite{REN2022102186}.
In the adaptive Kriging model of failure probability with respect to distribution parameters (FP-DP) \cite{PENG2022103353}, stopping is governed by two complementary rules: (i) a distance-based condition, where enrichment ceases once new samples are too close to existing ones, and (ii) an error-based condition, where termination occurs once the relative change in predicted failure probability between successive iterations drops below a small threshold.
In the improved active learning Kriging model with subset simulation (i-AKSS) \cite{ZHANG2022103256}, stopping criteria are defined by the maximum normalized prediction variability or by whether 99\% of points in the failure domain remain unchanged after 10 additional training points are added.
In the confidence interval squeezing (CIS) framework \cite{CHEN2022108639}, enrichment continues until the confidence interval width of the Kriging-estimated failure probability is sufficiently reduced, with global convergence ensured by bounding the relative error of the failure probability estimate.

Recent refinements include the $\epsilon$-stopping criterion in subdomain decomposition \cite{ZHANG2023109034}, error-based rules for sequential reliability updating \cite{FENG2023102366}, the global error-based learning function (GELF) \cite{ZHANG2024102464}, and the failure probability uncertainty-penalty function (FPUP) \cite{JIA2024102427}.
Other advances rely on probability stabilization \cite{KIM2024102404}, the most probable misclassification function in quantified active learning Kriging \cite{PRENTZAS2024103699}, or convergence criteria in domain-decomposition Monte Carlo schemes \cite{ZHANG2024109610}.
In addition, stopping rules have been extended to time- and space-dependent system reliability analysis, where termination is based on the maximum error in estimating the failure probability \cite{ZHAN2025110536}.

\textcolor{black}{Kriging-based stopping criteria have evolved from simple learning-function thresholds (such as the $U$- and EI-criteria) to more advanced error-based, stability-based, and hybrid formulations.
Learning-function-based approaches remain by far the most widely adopted, owing to their efficiency in concentrating samples near the limit state surface.
Their performance is, however, sensitive to surrogate misclassification, particularly in problems with multiple or disconnected failure regions, where premature convergence may occur.
In contrast, error- and confidence-based criteria provide more reliable control over estimation accuracy, but are typically more computationally demanding and are therefore less frequently used in practice.
More recent developments increasingly combine multiple criteria to balance accuracy, efficiency, and robustness, reflecting a broader shift toward hybrid stopping strategies. 
Overall, this progression highlights both the advancement of Kriging-based stopping methods and the continuing need for problem-specific adaptations in structural reliability analysis.
A further discussion of these criteria is presented in Sections \ref{subsection active learning} and \ref{subsection bayesian methods}, respectively, in the contexts of active learning and Bayesian reliability frameworks.}

\subsection{Response surfaces}
\label{subsection response surfaces}

Response surfaces aim to approximate the limit state function by constructing surrogate models as linear combinations of basis functions, such as polynomials, exponentials, radial basis functions, or splines.
Within this framework, stopping criteria play a central role in determining when adaptive refinement can be terminated, ensuring that the surrogate has reached an acceptable level of accuracy without expending unnecessary computational effort.
Because response surfaces are versatile yet often sensitive to model order and sampling strategy, the design of stopping rules has evolved in parallel with the development of different functional forms.

As for polynomial response surfaces, early studies \cite{BRANDYBERRY1990153, faravelli-1989-response-surface-approach-for-reliability-analysis, LUCIA1982281} emphasizes adequacy checks, where refinement halted once higher-order terms or additional sampling no longer improved predictive accuracy.
The sequential response surface method (SRSM) \cite{YINGWEILIU199439} further formalizes this idea, stopping when the difference between the true limit state function and its polynomial approximation at the updated center point falls within tolerance.
Later work adopts more pragmatic convergence measures \cite{GUAN2001429}, where iterations are terminated once the fitted polynomial produces stable estimates of failure probability.
More recently, the adaptive sparse response surface method \cite{GUIMARAES201812} combines stability of the reliability index with a bootstrap confidence interval check, ensuring both convergence and statistical robustness before stopping.

Exponential response surfaces have been proposed as alternatives when polynomial forms are inadequate. 
A dual criterion is adopted in \cite{HADIDI201715}, requiring convergence of both the reliability index and the limit state function at the estimated design point, reflecting a stricter view of model adequacy.
Similarly, in the hybrid approach \cite{10.1063/5.0074702}, the surrogate begins as a first-order polynomial but transitions to an exponential form in later iterations, with refinement halted once the reliability index stabilizes.
These examples highlight the use of stability-based rules that are tightly linked to the behavior of the reliability index, a key quantity in structural reliability.

Radial basis function surrogates bring additional flexibility but also require careful stopping rules.
In the reliability-based optimization of laminated composites \cite{GOMES2011186}, termination is governed not by the surrogate itself but by the genetic algorithm, with stopping based on population diversity or maximum generations.
In contrast, in \cite{LI201842}, the refinement process is terminated once the failure probability stabilizes, imposing both absolute and relative error tolerances.
Later contributions introduces more sophisticated criteria.
In the AK-ARBIS method \cite{YUN2020101891}, updates cease when the minimum value of the $U$-function is reached and the coefficient of variation of the failure probability falls below $0.05$, while $k$-fold cross-validation is employed in \cite{Shi2019}, halting once at least one surrogate produces a failure probability estimate within tolerance.
These approaches illustrate how error-based and variance-based rules can be tailored to the specific behavior of RBF surrogates.

Spline-based response surfaces offer yet another perspective.
In the successive response surface methodology (SRSM) \cite{4126283}, multivariate adaptive regression splines (MARS) are employed to refine search regions toward global optima, with stopping based on relative error or consistency across successive subregions.
More recently, in the adaptive isogeometric method \cite{BRACCO2024117287}, hierarchical $C^1$ splines are introduced over multi-patch domains approximate structural and shell problems.
Here, refinement terminated once a posteriori error estimators indicated that further updates would no longer significantly reduce discretization error, reflecting the integration of concepts in numerical analysis into stopping rules for response surface modeling.

\textcolor{black}{The strength of these stopping criteria lies in their ability to translate improvements in surrogate accuracy into tangible measures of convergence, such as the stabilization of the reliability index, diminishing prediction errors, or consistent performance across validation procedures.
In response surface-based approaches, stopping rules are typically driven by global approximation error or cross-validation measures, reflecting an emphasis on overall surrogate fidelity.
While these criteria are generally effective for problems with smooth and low-order nonlinear behavior, they may fail to capture localized features near the limit state surface.
Consequently, convergence in global error metrics does not necessarily guarantee accurate reliability estimation, particularly in rare-event or highly nonlinear scenarios.
In practice, the choice of a stopping criterion should be guided by both the surrogate formulation and the complexity of the problem.
Residual- or coefficient-based measures are often sufficient for low-order polynomial models \cite{VillanovaHyndmanSmithMilesPoli2013}, whereas confidence- or variance-based indicators are more suitable for nonlinear or high-dimensional settings \cite{Sacks1989}.
Future research may therefore benefit from adaptive stopping strategies that dynamically calibrate tolerance thresholds based on propagated uncertainty, while ensuring robustness against both overfitting and insufficient sampling.}

\subsection{Polynomial chaos expansions}
\label{subsection polynomial chaos expansions}

Polynomial chaos expansions (PCE) are widely employed in structural reliability analysis as surrogate models for computationally expensive functions, enabling efficient estimation of failure probabilities. Stopping criteria in PCE-based reliability analysis are essential to guarantee accuracy while avoiding unnecessary computational effort.
These criteria typically monitor the convergence of PCE coefficients, the error in failure probability estimates, or the stabilization of the reliability index.

Several representative strategies illustrate this principle.
In the resampling polynomial chaos method (RPCM) \cite{Notin01012010}, the polynomial expansion is iteratively enriched until the reliability index stabilizes, with stopping determined by the width of bootstrap confidence intervals.
Similarly, in the polynomial-chaos Kriging framework \cite{doi:10.1061/AJRUA6.0000870}, iterations halt once the relative difference between surrogate-based bounds and their mean estimate falls below a specified tolerance, thereby ensuring the accuracy of statistics such as failure probabilities or quantiles.
The active bootstrap-PCE (A-bPCE) algorithm \cite{MARELLI201867} defines convergence in terms of the variability of failure probability estimates across bootstrap replicates, requiring consistency of results over successive iterations.
In \cite{MODAK2024102467}, stopping criteria within the polynomial chaos expansion (PCE) framework are defined through a combination of convergence and confidence-based conditions.
Extensions of these ideas appear in ensemble approaches.
In \cite{ZHOU2022108778}, the stopping rule is based on the relative width of confidence bounds for the estimated failure probability, while in \cite{SONG2025120166}, the PC-Kriging surrogate is updated until successive estimates of the augmented failure probability (AFP) differ by less than a prescribed threshold.
In the PCE-based multi-objective reliability framework (SEP-MRS) \cite{TENG2025111230}, its stopping criterion relies on cross-validation, monitoring the relative improvement of prediction error during iterative hyperparameter optimization.

\textcolor{black}{Broadly speaking, PCE-based stopping strategies prove effective because they assess how reliably the expansion predicts system failure, moving beyond mere coefficient convergence (for instance, \cite{doi:10.1061/AJRUA6.0000870}).
These approaches commonly rely on variance-based or spectral error indicators, leveraging the orthogonal structure of the expansion to provide clear convergence diagnostics.
When combined with quantitative measures such as confidence bounds or the stabilization of failure probability estimates, they offer a practical and statistically grounded basis for termination.
Their effectiveness is, however, typically limited to problems with smooth responses and well-behaved input uncertainties.
In the presence of strong nonlinearity or discontinuities in the limit state function, these criteria may become less reliable, as truncation of higher-order terms can lead to underestimation of tail behavior and reduced accuracy in rare-event regimes.
In practice, achieving a balance between accuracy and computational efficiency requires selecting criteria consistent with the problem structure.
Future research may therefore benefit from adaptive or hybrid stopping strategies that integrate both statistical and structural indicators, enabling more robust convergence assessment in complex applications.}

\subsection{Support vector machines}
\label{subsection support vector machines}

In structural reliability analysis, support vector machines (SVMs) have gained prominence as surrogate models capable of approximating complex limit state functions with relatively small training datasets \cite{ROY2023109126}.
As with other surrogate modeling frameworks, a central challenge lies in formulating effective stopping criteria that guarantee sufficient accuracy in the failure domain approximation while keeping computational demands under control.

Several strategies have been proposed to address this issue.
In the adaptive SVM Monte Carlo simulation (ASVM-MCS) method \cite{PAN201785}, training set enrichment continues until the proportion of Monte Carlo samples lying inside the SVM margin, along with its exponential fit, falls below a target threshold, and the slope of the fitted curve is sufficiently small.
Termination is further guided by the coefficient of variation of the estimated failure probability, ensuring both accuracy and stability.
A related approach is presented in \cite{ROY2020106948}, where support vector regression (SVR) is embedded within a two-stage adaptive sampling scheme.
Here, the design of experiments is sequentially enriched near the estimated failure surface, and the process terminates once the change in the failure probability estimate between successive iterations falls below a tolerance.

Hybrid and error-based stopping rules have also been explored.
In adaptive Bayesian SVR (ABSVR) \cite{WANG2021114172}, a hybrid criterion combines bootstrap confidence estimation with a stabilization check on reliability indices across consecutive iterations.
The active learning loop ends once either the confidence interval narrows sufficiently or the reliability index stabilizes, avoiding premature termination while preventing excessive enrichment.
Similarly, an inner-loop rule is employed in \cite{PEPPER2022108635} to halt enrichment based on the difference between the empirical failure probability and its conservative upper bound, ensuring that the SVM surrogate stabilizes before the outer loop terminates.

Other works embed SVR within broader ensemble (Section \ref{subsection ensemble models}) or time-dependent frameworks.
In the AEM-PDEM approach \cite{ZHOU2022108778}, regression-type SVM predictions are governed by ensemble-level rules, terminating once confidence-bound-based estimates of failure probability converge across consecutive iterations.
The EMSVR-MCS method \cite{https://doi.org/10.1002/qre.3747} also adopts SVR in an ensemble setting, where convergence is declared when the failure probability estimate stabilizes across four successive iterations.
Extending to time-variant problems, TASVM-MCS is proposed in \cite{ZHENG2025110510}, where adaptive enrichment stops once the maximum variation of the time-dependent failure probability function (TFPF) across iterations falls below a prescribed threshold, ensuring convergence across all time instants.

\textcolor{black}{The effectiveness of stopping criteria in SVM-based reliability methods arises from jointly monitoring classifier stability and the accuracy of reliability estimates, translating variations in the margin region, prediction error, or failure probability into quantitative convergence indicators.
These approaches, typically based on margin evolution, classification stability, or hybrid error measures, exploit structural properties of the model (notably the margin and kernel) to detect when further enrichment yields negligible improvement near the decision boundary, and are particularly effective in high-dimensional settings.
Their performance may, however, be sensitive to the distribution of training data and poorly sampled failure regions.
In particular, stability-based criteria can lead to premature convergence when the learned boundary stabilizes locally without accurately representing the true failure domain.
Although not explicitly formalized in the literature, margin- or probability-based rules appear well suited to classification-oriented problems, whereas confidence- or variance-based measures are more appropriate for regression or ensemble frameworks \cite{ROY2023109126}.
Developing adaptive stopping schemes that couple margin evolution with posterior uncertainty and convergence diagnostics remains a promising direction for improving robustness and efficiency in both static and time-variant (Section \ref{subsection time-variant reliability analysis}) reliability analyses.}

\subsection{Neural networks}
\label{subsection neural networks}

Artificial neural networks (ANNs) and deep neural networks (DNNs) have been widely used to construct surrogate models of computationally intensive limit state functions \cite{Afshari_2023, Chojaczyk_Teixeira_Neves_Cardoso_GuedesSoares_2015}.
Stopping criteria play a dual role: they regulate surrogate training to prevent overfitting and determine when iterative simulations or optimization can terminate once predictive accuracy or estimator convergence is adequate.

In early applications, such as \cite{GOMES2011186}, no intrinsic stopping criterion is introduced for ANN training.
Termination is set by the surrounding optimization framework (for instance, population standard deviation or maximum generations).
The ANN itself was trained offline with a fixed dataset.
A more conventional ML strategy is seen in \cite{ZHANG20181}, where learning stops once the network error falls below a tolerance or an iteration cap is reached.

Later developments move stopping criteria from ANN training to active learning loops.
For instance, in subset simulation (Section \ref{subsection subset simulation}) with surrogate assistance \cite{CUI201912}, network updates stop when a predefined condition for the outer loop is met.
Similarly, in ensemble approaches \cite{REN2022102186}, termination occurs either when the evaluation budget is exhausted or when ANN- and Kriging-based estimates converge, ensuring efficiency and stability.

More recently, DNNs have gained prominence in structural reliability \cite{10.1115/1.4047636}.
In \cite{https://doi-org.utokyo.idm.oclc.org/10.1111/mice.12359}, training relies on minimizing Euclidean mean squared error for regression and binary cross-entropy for classification, with standard backpropagation until the loss is sufficiently reduced.
In the adaptive DNN surrogate of \cite{LIEU2022116104}, stopping is based on two rules.
Specifically, training ends after 20 iterations or when relative and absolute errors of failure probability fall below set tolerances.
Additionally, once the sample size exceeds a threshold, noisy or uninformative points are discarded so training concentrates near the limit state surface.
In \cite{GARG2022114811}, DeepONet training minimizes mean squared error through iterative backpropagation over a fixed number of epochs, with no explicit early-stopping rule such as validation error thresholds.
In the adaptive spatiotemporal sampling physics-informed neural network (ASTS-PINN) \cite{HU2026111560}, the iterative training is halted once the loss function shows stagnation.

\textcolor{black}{A typical challenge in ANN- and DNN-based surrogate modeling is the risk of stopping either too early or too late, often driven by unstable or noisy validation performance and overfitting, especially when training data are sparse or imbalanced \cite{Afshari_2023}.
Accordingly, stopping criteria must balance predictive fidelity with computational budget, ensuring adequate learning while avoiding overfitting \cite{Chojaczyk_Teixeira_Neves_Cardoso_GuedesSoares_2015}.
In practice, neural network-based approaches rely primarily on loss-based or validation-error-based criteria, which are straightforward to implement within standard training frameworks.
These indicators are, however, only indirectly related to reliability accuracy and may not adequately reflect errors in failure probability estimation, thereby requiring additional problem-specific measures to ensure consistency between training convergence and reliability predictions.
The formulation of stopping rules should therefore account for both network architecture and data characteristics.
While fixed iteration limits or error thresholds may suffice for low-dimensional problems, adaptive or hybrid schemes are generally necessary in active learning and rare-event settings.
Future research may focus on uncertainty-informed mechanisms that dynamically adjust tolerance thresholds and promote stable convergence across coupled surrogates, while incorporating advances from related methodological domains.}

\subsection{Multifidelity models}
\label{subsection multifidelity models}

In structural reliability analysis, multifidelity models aim to combine information sources of varying accuracy and cost to efficiently estimate failure probabilities.
Adaptive strategies, often within active learning frameworks (Section \ref{subsection active learning}), guide both the selection of fidelity levels and sampling locations, typically emphasizing regions near the limit state surface or areas of high predictive uncertainty.
Because high-fidelity simulations are computationally expensive, stopping criteria play a central role in balancing predictive accuracy against computational cost.
Unlike single-fidelity approaches, multifidelity frameworks require stopping rules that reflect not only surrogate convergence but also the allocation of computational resources across fidelity levels.

A wide range of stopping strategies has been proposed in this context.
Error-based criteria are particularly prominent.
For instance, in the AMK-MCS+AEFF method \cite{yi_active-learning_2021}, the adaptive process terminates once the relative error between true and surrogate-predicted failure probabilities falls below a prescribed tolerance, ensuring convergence without excessive high-fidelity evaluations.
Similarly, in \cite{YI2022108247}, the bootstrap stopping criterion (BSC) of \cite{yi_zhou_cheng_liu_2020} is extended to multifidelity Kriging, yielding the MF-BSC-Believer method, where termination is guided by bootstrap-based confidence estimates combined with adaptive fidelity selection.

Alternative approaches rely on stability measures and learning functions.
In the AMGPRA framework \cite{ZHANG2022102141}, stopping is driven by the collective learning function (CLF), with enrichment halted once its maximum value falls below a threshold, supplemented by the monitoring of the coefficient of variation.
In bi-fidelity neural network approaches \cite{LIMA2023109492} (Section \ref{subsection neural networks}), training is terminated when the average relative error across fidelity levels satisfies a tolerance or when a maximum number of epochs is reached.
In addition, methods combining multifidelity surrogates with subset simulation \cite{NADERI2026111739} employ adaptive sampling strategies that terminate when the coefficient of variation of the failure probability estimate reaches a target level.

More recent developments explicitly integrate multiple indicators within unified stopping strategies.
In the multifidelity co-Kriging Monte Carlo simulation framework (AMCK-MCS) \cite{wanglinjun2025}, termination is based on a combination of confidence interval estimation, relative error measures, and uncertainty weighting.
In hierarchical Kriging-based active learning \cite{FENG2026111759}, a two-stage convergence framework is adopted, where the low-fidelity surrogate is first refined until global changes stabilize, followed by a high-fidelity stage governed by error-based stopping criteria such as the global rate of misclassification.
These hybrid approaches aim to ensure that high-fidelity evaluations are performed only when they significantly reduce overall uncertainty.

\color{black}
Despite these advances, several fundamental challenges arise when interpreting stopping criteria in multifidelity settings. 
A key issue is the implicit assumption that information obtained from low-fidelity (LF) models can be directly transferred to high-fidelity (HF) responses.
While often adopted in practice, this assumption may not hold in the presence of structural discrepancies or complex model-form differences.
In many cases, stopping rules quantify uncertainty within a surrogate hierarchy without rigorously verifying whether convergence observed at the LF level reflects true HF accuracy. 
Furthermore, such strategies often rely on the assumption that LF-HF mappings are smooth or well behaved, whereas in practice they may exhibit strong nonlinearities or discontinuities, limiting the reliability of convergence indicators derived from LF behavior.

From a broader perspective, convergence in multifidelity frameworks is not established through strict analytical guarantees but rather inferred through surrogate approximations and empirical correlations.
As a result, stopping criteria in these settings should be interpreted as providing statistical confidence under modeling assumptions, rather than definitive guarantees of accuracy.
This limitation becomes particularly important in problems involving weak cross-fidelity correlation, pronounced model discrepancy, or highly nonlinear failure domains.

In summary, multifidelity stopping strategies typically rely on error-, variance-, and confidence-based criteria to balance predictive accuracy and computational cost.
Their practical effectiveness depends critically on the consistency between fidelity levels and the adequacy of the surrogate representation.
Consequently, no single class of stopping criteria is universally reliable, and robust performance is generally achieved through hybrid approaches that combine multiple indicators while accounting for uncertainty propagation and model discrepancy.
Future research may therefore focus on developing uncertainty-aware stopping rules that jointly manage sampling and fidelity allocation, integrate real-time cost-benefit considerations, and extend to multifidelity ensembles and time-variant reliability frameworks (Section \ref{subsection time-variant reliability analysis}).

\color{black}
\subsection{Reduced-order models}
\label{subsection reduced-order models}

In structural reliability analysis, reduced-order models (ROMs) and low-rank approximations (LRAs) are widely employed to accelerate the evaluation of expensive high-fidelity models (Section \ref{subsection multifidelity models}).
Stopping criteria are essential in this context, ensuring that model refinement halts once sufficient accuracy is achieved while preventing unnecessary enrichment.
Unlike single-fidelity settings, ROM-based approaches often rely on convergence or stability measures that reflect both surrogate accuracy and statistical reliability of failure probability estimates.

Several contributions implement error- or variance-based rules. 
In the singular value decomposition-based ROM for severe accident analysis codes \cite{Matsushita03052020}, adaptive sampling continues until convergence of the estimated core damage boundary is observed, and only singular vectors covering more than 99.99\% of the variance are retained.
Similarly, in the proper generalized decomposition (PGD)-based adaptive importance sampling framework \cite{robens-radermacher}, surrogate refinement ceases once the reliability index stabilizes, while the sampling stage terminates when the variance of the failure probability estimate drops below a tolerance.

Stability criteria have also been emphasized in spectral and ensemble-based ROM frameworks.
In the stochastic spectral embedding-based reliability (SSER) method \cite{WAGNER2022102179}, local spectral expansions are iteratively enriched until the bounds of the generalized reliability index converge, with termination declared when their relative difference remains below 3\% over three consecutive iterations, supplemented by a computational budget cap.
In the AEM-PDEM framework \cite{ZHOU2022108778}, ROMs such as low-rank approximations are embedded in an ensemble (Section \ref{subsection ensemble models}), where stopping is determined by the stabilization of failure probability confidence bounds across consecutive iterations.

\textcolor{black}{Broadly speaking, the effectiveness of stopping criteria in reduced-order modeling arises from monitoring convergence through indicators such as stabilization of the reliability index (Section \ref{section approximate methods}), reduction of approximation error, or contraction of confidence bounds.
In practice, they are typically based on residual or approximation error measures, which provide efficient control of model reduction accuracy.
A common pitfall, however, is premature termination, which may occur when variance-based measures underestimate epistemic uncertainty, when truncation eliminates important modes, or when reduced bases fail to capture localized nonlinearities or disjoint limit states \cite{WAGNER2022102179}.
As a consequence, these criteria may underestimate the impact of model reduction on reliability predictions, particularly when critical failure modes are insufficiently represented.
Their effectiveness therefore depends on the alignment between reduction error and reliability-sensitive features.
The choice of a stopping rule should be consistent with the reduction strategy: variance- or residual-based measures are generally suited to projection or decomposition methods (such as singular value decomposition and proper generalized decomposition), whereas stability- or confidence-based criteria align more naturally with ensemble or stochastic spectral ROMs.
Advancing multi-criteria stopping schemes with probabilistic error control remains a promising direction, particularly for multifidelity (Section \ref{subsection multifidelity models}), time-variant (Section \ref{subsection time-variant reliability analysis}), and data-driven ROMs.}

\subsection{Ensemble models}
\label{subsection ensemble models}

Ensemble models have proven highly effective in advancing structural reliability analysis, aiming to construct accurate surrogates for complex limit state functions while enhancing the estimation of failure probabilities.
A central challenge in their adaptive development lies in formulating effective stopping criteria, which determine when to halt refinement and finalize the ensemble.

Several recent contributions illustrate how stopping conditions are tailored to ensemble modeling.
In the AEM-PDEM framework \cite{ZHOU2022108778}, ensembles of polynomial chaos Kriging (Section \ref{subsection polynomial chaos expansions}), support vector regression (Section \ref{subsection support vector machines}), and low-rank approximation (Section \ref{subsection reduced-order models}) are iteratively refined until the relative difference between upper and lower bounds of the estimated failure probability remains within a prescribed tolerance for consecutive iterations.
This criterion provides robustness across different metamodel types.
In \cite{REN2022102186}, ensembles that combine ANNs (Section \ref{subsection neural networks}) and Kriging are governed by two complementary rules.
In particular, a budget-based criterion that limits the number of expensive model evaluations, and a stability-based criterion that halts learning once the failure probability estimate of the ensemble agrees closely with those from individual surrogates, typically within 1\%.
Similarly, the adaptive stochastic configuration network (SCN) ensemble approach \cite{LIU2024121633} integrates multiple SCN base learners, with sampling continuing until either a preset maximum number of new samples is reached or the failure probability estimate of the ensemble stabilizes across iterations.

Other ensemble frameworks emphasize stability-based termination.
In \cite{SONG2025120166}, an adaptive ensemble integrating Kriging and polynomial chaos Kriging surrogates (Section \ref{subsection polynomial chaos expansions}) refines its predictions through a learning function, with convergence declared once the change in augmented failure probability estimates between successive iterations falls below a specified threshold.
This ensures accurate estimation of the failure probability function in reliability-based design optimization (Section \ref{subsection reliability-based design optimization}), while avoiding excessive model enrichment.
Likewise, the EMSVR-MCS method \cite{https://doi.org/10.1002/qre.3747} combines ensembles of SVR metamodels (Section \ref{subsection support vector machines}) with Monte Carlo simulation, and defines convergence as stability of the ensemble-estimated failure probability across the last four iterations, providing a statistically grounded stopping rule.

\textcolor{black}{In ensemble models, the effectiveness of stopping criteria hinges on assessing convergence at the aggregate level, where the diversity and mutual compensation of constituent surrogates can be exploited.
Accordingly, these approaches often rely on agreement-based or variance-based indicators, leveraging the diversity of model predictions to assess convergence and improve robustness against individual model biases.
Typical pitfalls, however, arise from inter-model correlation or hidden local inaccuracies near the limit-state surface \cite{ZHOU2022108605}, which may lead to misleading convergence signals when multiple models share similar structural errors. 
Consequently, ensemble agreement does not necessarily guarantee accuracy in reliability estimation, particularly in complex or extrapolatory regimes.
In practice, the choice of a stopping criterion depends on the ensemble composition and computational budget.
In general, stability-based thresholds perform well for heterogeneous ensembles, whereas confidence- or budget-based rules are more suitable for expensive simulations.
Future research may focus on multilevel stopping strategies that incorporate diversity metrics, posterior uncertainty, and dynamic weighting to promote more robust and interpretable ensemble frameworks.}

\color{black}
\subsection{Synthesis and practical guidance}
\label{subsection summary 5}

Before concluding this section, we synthesize the main cross-framework trends and organize them into a comparative overview of stopping strategies, as summarized below in Table \ref{table:stopping_summary_frameworks}.
The table provides a unified perspective on the dominant criteria, strengths, and limitations across advanced modeling frameworks, and can be interpreted in relation to the performance dimensions outlined in Table \ref{table:performance_metrics}.
In particular, it highlights how different stopping strategies reflect trade-offs among accuracy, computational cost, and robustness, which serve as common benchmarks for evaluating their practical effectiveness across the literature.

\begin{table}[!ht]\color{black}
\centering
\caption{Comparison of dominant stopping criteria, strengths, and limitations across advanced modeling frameworks}

\vspace{0.5em}
\begin{tabular}{p{0.16\columnwidth} | p{0.22\columnwidth} | p{0.24\columnwidth} | p{0.26\columnwidth}}
\hline
\textbf{Framework} & \textbf{Dominant stopping criteria} & \textbf{Strengths} & \textbf{Limitations / failure modes} \\
\hline

Kriging models 
& Learning-function-based, confidence-based, error-based 
& Efficient adaptive sampling near the limit state; strong empirical performance in active learning 
& Sensitive to surrogate misclassification; may miss disconnected failure regions; reliance on heuristic thresholds \\

\hline

Response surfaces and polynomial chaos expansions 
& Error-based (cross-validation), variance-based 
& Direct control of approximation error; interpretable convergence 
& May underrepresent rare-event regions; global error does not ensure local accuracy near the failure boundary \\

\hline

Support vector machines
& Margin-based, stability-based, hybrid error-based 
& Effective boundary classification; good performance in high-dimensional spaces 
& Sensitive to training data distribution; may stabilize prematurely around incorrect boundaries \\

\hline

Neural networks
& Loss-based, validation error, epoch-based 
& Easy to implement; compatible with gradient-based optimization 
& Loss convergence does not ensure reliability accuracy; weak direct linkage to failure probability estimation \\

\hline

Multifidelity models 
& Error-based, CoV-based, confidence-based 
& Efficient balance between accuracy and computational cost 
& Assumes LF-HF consistency; LF convergence may not imply HF accuracy; model discrepancy issues \\

\hline

Reduced-order models 
& Error-based, residual-based 
& Reduced computational cost; suitable for complex physical systems 
& Model reduction error may dominate; stopping criteria may underestimate true error \\

\hline

Ensemble models 
& Agreement-based, variance-based, hybrid 
& Improved robustness through model diversity; enables uncertainty quantification 
& May give false confidence if models share bias; aggregation can mask individual model failure \\
\hline
\end{tabular}
\label{table:stopping_summary_frameworks}
\end{table}

Building on the comparative overview in Table \ref{table:stopping_summary_frameworks}, several consistent patterns emerge across advanced modeling frameworks.
Learning-function- and confidence-based criteria dominate in surrogate-based approaches, particularly in Kriging and active learning, due to their efficiency in concentrating sampling near the limit state surface.
These criteria, however, remain sensitive to surrogate misclassification and may lead to premature convergence in complex or multimodal failure domains.
Error- and variance-based criteria provide more direct control over estimation accuracy, but are often computationally demanding and therefore less frequently employed in high-cost settings.
Meanwhile, stability-based criteria are widely used due to their simplicity, yet they tend to be less reliable when employed in isolation, as they may fail to detect global inaccuracies.
Overall, no single class of stopping criteria consistently outperforms others across frameworks, highlighting a fundamental trade-off between accuracy, robustness, and computational efficiency.
As a result, hybrid strategies combining multiple indicators have emerged as the most robust practical solutions.

From a practical standpoint, several guidelines emerge.
For surrogate-based and active learning methods, learning-function-based criteria are generally preferred, provided that surrogate accuracy near the limit state surface is carefully controlled.
For Monte Carlo-based approaches, variance- or coefficient-of-variation-based criteria remain the most reliable choices when computational cost is manageable.
Stability-based criteria should not be used in isolation, but can serve as complementary indicators when combined with error- or variance-based measures.
In complex or rare-event settings, hybrid strategies combining multiple indicators offer the most robust performance across problem types.





\color{black}

\section{Stopping criteria for advanced methodological frameworks}
\label{section advanced methodological frameworks}

Building on the preceding overview of stopping criteria in advanced modeling frameworks (Section \ref{section advanced modeling frameworks}), we now turn to their treatment within advanced methodological approaches, namely active learning reliability methods,
Bayesian methods,
adaptive response surface methods,
reliability-based design optimization,
and time-variant reliability analysis,
as respectively presented in Sections \ref{subsection active learning}-\ref{subsection time-variant reliability analysis}.
In particular, for clarity of presentation from a methodological perspective, we organize the subsequent discussion of active learning reliability methods (Section \ref{subsection active learning}) and Bayesian methods (Section \ref{subsection bayesian methods}) around categories of stopping criteria, such as learning functions, error-based rules, and stability-based approaches, while avoiding excessive technical detail so as to highlight the broader landscape of stopping criteria across diverse frameworks.
As noted earlier, many references have already appeared, sometimes multiple times, in Section \ref{section advanced modeling frameworks} and reappear in the upcoming section, reflecting the frequent integration of multiple models and methodologies.

\subsection{Active learning reliability methods}
\label{subsection active learning}

Active learning reliability (ALR) methods have seen rapid development over the past decade, thanks to their ability to provide accurate and computationally efficient estimates of small failure probabilities in structural reliability analysis \cite{MOUSTAPHA2022102174, TEIXEIRA2021102019}.
These methods iteratively refine surrogate models, most commonly based on Gaussian process (GP) modeling or Kriging (Section \ref{subsection kriging models}), by selectively enriching the training set with informative samples.
In line with growing interest in stopping criteria in this context, several reviews have been presented in the literature (for instance, \cite[Table 6]{ALLAHVIRDIZADEH2023109406}, \cite[Table 1]{DANG2025102539} and
\cite[Table 1]{Shi2019}, to name just a few).
Here, we start with three types of criteria employed in the relevant literature, based on learning functions (Section \ref{subsubsection learning functions}), errors (Section \ref{subsubsection ALR error}) and stability (Section \ref{subsubsection ALR stability}), followed by further related works and outlook (Section \ref{subsubsection further related works}).

\subsubsection{Learning functions}
\label{subsubsection learning functions}

In active learning reliability methods, the enrichment phase proceeds by selecting at least one additional sample point as long as convergence has not yet been attained.
This selection is typically guided by a learning function, which identifies candidate points expected to improve the surrogate model, and consequently the accuracy of the estimated failure probability, when incorporated into the design.
\textcolor{black}{In the literature, learning functions are often distinguished from stopping criteria, the latter being associated with explicit convergence conditions \cite{MOUSTAPHA2022102174}. However, building upon \cite[Section 2]{SHI2020106857}, they may also be interpreted as implicit stopping mechanisms within iterative enrichment frameworks. More precisely, learning functions act as convergence indicators rather than termination rules in themselves, although they are frequently combined with explicit stopping conditions in practice.}

First, with a realization of the Gaussian process $G({\bf x})\sim \mathcal{N}(\mu_{G}({\bf x}),\sigma^2_G({\bf x}))$, efficient global reliability analysis (EGRA) \cite{bichon_efficient_2008} is designed to stop the enrichment scheme when the value of the expected feasibility function (EFF), defined as 
\begin{align*}
    {\rm EFF}({\bf x})&:= \mu_{G}({\bf x}) \left[ 2 \Phi \left( - \frac{\mu_{G}({\bf x})}{\sigma_G (\mathbf{x})} \right) - \Phi \left( \frac{-\epsilon - \mu_{G}({\bf x})}{\sigma_G (\mathbf{x})} \right) - \Phi \left( \frac{\epsilon - \mu_{G}({\bf x})}{\sigma_G (\mathbf{x})} \right) \right] \\
    &\qquad - \sigma_{G} (\mathbf{x}) \left[ 2 \phi \left( - \frac{\mu_{G}({\bf x})}{\sigma_G (\mathbf{x})} \right) -\phi \left( \frac{-\epsilon - \mu_{G}({\bf x})}{\sigma_G (\mathbf{x})} \right) - \phi \left( \frac{\epsilon - \mu_{G}({\bf x})}{\sigma_G (\mathbf{x})} \right) \right] \\
    &\qquad + \epsilon\left[ \Phi \left( \frac{\epsilon - \mu_{G}({\bf x})}{\sigma_{G} (\mathbf{x})} \right) - \Phi \left( \frac{-\epsilon - \mu_{G}({\bf x})}{\sigma_G (\mathbf{x})} \right) \right],
\end{align*}
is lower than $10^{-3}$, indicating convergence of the GP model in the region near the limit state surface (see also \cite{WEN2016170, zhang_reif_2019}).
Similarly, adaptive Kriging Monte Carlo simulation (AK-MCS) \cite{ECHARD2011145} also exemplifies this paradigm, where the iteration is stopped when the $U$-function \cite{CADINI2020101998, ECHARD2013232, HUANG201686, ma_adaptive_2022}, defined as 
\[ 
 U({\bf x}):=\frac{|\mu_{G}({\bf x})|}{\sigma_{G}({\bf x})},
\]
is greater than $2$ for all candidate points.
This condition ensures a misclassification probability below $2.28\%$ for all samples considered in estimating the failure probability.
In practice, it is highly conservative and tends to introduce an excessive number of points.
Some studies have tried relaxing this conservative condition, either by considering the whole candidate set through an average (for instance, \cite{JIAN2017494, LELIEVRE20181, SUN2017152}), or by considering convergence when only a small proportion of the candidate set does not comply with the conservative condition of the $U$ function being greater than $2$ for all candidate points \cite{FAURIAT2014137, Moustapha2016}.
We note that the (absolute) CoV, defined as the reciprocal of (and thus equivalent to) the $U$-function, is adopted in \cite{ZHANG2022103256}.

In the literature, more advanced alternatives have also been proposed, like the $H$-function \cite{lv_new_2015}:
\begin{multline*}
    H({\bf x}):= \Bigg|\ln \left(\sqrt{2\pi}\sigma_ {G} (\mathbf{x})+1/2\right) \left[ \Phi \left( \frac{2 \sigma_ {G} (\mathbf{x})-\mu_{G} (\mathbf{x})}{\sigma_ {G} (\mathbf{x})} \right) - \Phi \left( \frac{-2 \sigma_ {G} (\mathbf{x})-\mu_{G} (\mathbf{x})}{\sigma_ {G} (\mathbf{x})} \right) \right] \\
    - \left[ \frac{2\sigma_{G} (\mathbf{x})-\mu_{G}(\mathbf{x})}{2} \phi \left(\frac{2\sigma_{G} (\mathbf{x})-\mu_{G}(\mathbf{x})}{\sigma_ {G} (\mathbf{x})} \right)+\frac{2\sigma_{G} (\mathbf{x})+\mu_{G}(\mathbf{x})}{2} \phi \left(\frac{-2\sigma_{G} (\mathbf{x})-\mu_{G}(\mathbf{x})}{\sigma_ {G} (\mathbf{x})} \right) \right] \Bigg|,
\end{multline*}
the reliability-based expected improvement function (REIF) \cite{jones_schonlau_welch_1998, zhang_reif_2019}:
\[
 {\rm REIF}({\bf x}):=\mu_{G}(\mathbf{x})\left(1-2\Phi\left(\frac{\mu_{G}(\mathbf{x})}{\sigma_{G}(\mathbf{x})}\right)\right)+\sigma_{G}(\mathbf{x})\left(w-\sqrt{\frac{2}{\pi}}\exp\left[\frac{1}{2}\frac{\mu^2_{G}(\mathbf{x})}{\sigma^2_{G}(\mathbf{x})}\right]\right),
\]
for some adjustment constant $w$, the potential risk function (PRF) \cite{HONG2022114835}:
\[
{\rm PRF}({\bf x}):=\frac{\min_k\|{\bf x}_k-{\bf x}\|}{\max_{k_1,k_2}\|{\bf x}_{k_1}-{\bf x}_{k_2}\|}\exp\left[-c \frac{|G({\bf x})|}{\max_k g({\bf x}_k)}\right],
\]
for some positive $c$, the most probable misclassification function \cite{PRENTZAS2024103699}: 
\[
\pi_q({\bf x}):=2\min\{\pi({\bf x}),1-\pi({\bf x})\}, \quad \pi({\bf x}):=\mathbb{P}(G({\bf x})\le 0)=\Phi\left(\frac{0-\mu_{G}({\bf x})}{\sigma_{G}({\bf x})}\right),
\]
the least improvement function (LIF) and uncertainty function (UF) \cite{SUN2017152}, respectively, defined as
\[
 {\rm LIF}({\bf x}):=\Phi(-U({\bf x}))f({\bf x})\mathbb{E}[|G({\bf x})|^d],\quad {\rm UF}:=\int_{\Omega}\Phi(-U({\bf x}))f({\bf x})d{\bf x},
\]
the failure probability uncertainty-based learning function (FPU) \cite{JIA2024102427}:
\[
 {\rm FPU}({\bf x}):=\sqrt{\Phi\left(\frac{-\mu_G({\bf x})}{\sigma_G({\bf x})}\right)\Phi\left(\frac{\mu_G({\bf x})}{\sigma_G({\bf x})}\right)}\frac{f({\bf x})}{w({\bf x})},
\]
a generalized $U$-function \cite{Persoons2023},
\[
  L^p({\bf x}):=\Phi\left(\frac{-|\mu_G({\bf x})|}{(\sigma_G({\bf x}))^p}\right)\frac{f({\bf x})}{w({\bf x})},
\]
where $w({\bf x})$ denotes an importance sampling density, the expected risk function (ERF) \cite{yang_liu_gau_zhang_gao_2015, YANG20153954}, defined as
\[
 {\rm ERF}({\bf x}):=-{\rm sgn}(\mu_{G}({\bf x}))\mu_{G}({\bf x})\Phi\left(-{\rm sgn}(\mu_{G}({\bf x}))\frac{\mu_{G}({\bf x})}{\sigma_{G}({\bf x})}\right)+\sigma_{G}({\bf x})\phi\left(\frac{\mu_{G}({\bf x})}{\sigma_{G}({\bf x})}\right),
\]
along with many variants, including the folded normal based expected improvement function (FNEIF) \cite{SHI2020106857}, the collective learning function (CLF) \cite{ZHANG2022102141}, reliability-based lower confidence bounding learning function \cite{yi_zhou_cheng_liu_2020}, relative entropy function (REF) \cite{dpr.2023.18}, sign loss function \cite{ZHENG2026111493}, sample-based expected uncertainty reduction (SEUR) \cite{PENG2024102384}, a modified uncertainty function for multiple simulation models \cite{YANG2024102426}, bootstrap polynomial chaos expansion methods \cite{MARELLI201867, MODAK2024102467}, a gradient-assisted learning function \cite{NAN2024109944}, multiple response handling \cite{ma_adaptive_2022}, region-adaptive or parallelized schemes \cite{WEN2016170}, as well as the expected information gain (EIG) \cite{LI2025103850}, which, while not a learning function in the strict sense, serves an equivalent purpose.

We stress that the above list is by no means exhaustive.
Moreover, even further lines of learning functions are expected to emerge, particularly as advances in machine learning (Section \ref{subsection neural networks}) and multifidelity modeling (Section \ref{subsection multifidelity models}) continue to inspire novel strategies tailored to reliability analysis.

The choice of a learning function plays a pivotal role in active learning reliability methods.
The $U$-function continues to serve as a widely adopted reference measure, valued for its simplicity and its clear focus on areas of greatest classification ambiguity near the limit state surface.
It has further been employed in the context of evaluating the global reliability sensitivity index \cite{YUN2025103848, yun2024}.
The Expected Feasibility Function (EFF) offers a broader formulation by accounting for both the distance to the limit state and the local predictive variance, thereby providing improved robustness in highly nonlinear problems.
Beyond these, numerous alternative learning functions have been proposed, as described above, to enhance sampling efficiency under reliability scenarios.
In some instances (such as \cite{DANG2025118092, YANG2024102426}), multiple learning functions are combined to leverage their complementary strengths.
In general, the selection of a learning function should depend on the problem characteristics, including dimensionality, failure probability level, surrogate model type, and computational budget, often with simpler functions preferred in low-dimensional or well-behaved cases, and more sophisticated criteria proving advantageous in high-dimensional settings.

\color{black}
In practical engineering applications, $U$-function-based criteria remain the most widely adopted due to their simplicity and computational efficiency, whereas EFF- and uncertainty-based criteria are generally preferred in highly nonlinear problems where classification confidence alone may be insufficient. More sophisticated learning functions can provide additional robustness, although often at the cost of increased computational complexity and parameter tuning requirements.
In general, the selection of a learning function should depend on the problem characteristics, including dimensionality, failure probability level, surrogate model type, and computational budget.

\color{black}
\subsubsection{Error-based stopping criteria}
\label{subsubsection ALR error}

The second class of stopping criteria is built upon the accuracy of the failure probability or surrogate predictions.
In \cite{DUBOURG201347}, a leave-one-out estimate of a correlation factor is used, with refinement halted once this factor reaches the order of magnitude of one.
In \cite{CHEN2022108639, DUBOURG2011, ZHOU2024103580}, stopping rules are formulated from the confidence interval of the failure probability.
For surrogate models without built-in error measures, analogous bounds have been derived using cross-validation \cite{Shi2019} or bootstrap replicates \cite{MARELLI201867}.
In ensemble-based reliability analysis (Section \ref{subsection ensemble models}), relative error of the failure probability forms the basis of the criterion \cite{ZHOU2022108778}, while \cite{ELHAJ2021102011, HONG2025102546, KIM2024102404} employ relative errors of estimated probabilities more generally.
In \cite{LIEU2019599}, the relative error of predicted responses at selected sample points defines the stopping rule within reliability-based design optimization (Section \ref{subsection reliability-based design optimization}).
In two-stage active learning Kriging (AK-TSESC) framework \cite{ZHAO2026111603}, that integrates global exploration and local exploitation, the stopping rule is rigorously formulated using the maximum relative error of failure probability estimates, ensuring convergence is neither prematurely declared nor excessively delayed.
For high-dimensional low-failure probability problems \cite{RAJAK2024103662}, enrichment is terminated once three consecutive estimates fall below a prespecified value.
Collectively, these statistical indicators provide quantitative and practical bases for terminating simulations once sufficient confidence in the failure probability estimate has been secured.

These error-based stopping criteria work well thanks to their statistical grounds, ensuring consistent performance across sampling, surrogate, and  ensemble  frameworks, while remaining adaptable to different levels of model uncertainty.
In practice, selecting an appropriate criterion involves balancing theoretical soundness, computational cost, and sensitivity to uncertainty, often favoring indicators that align naturally with the chosen learning function or reliability estimator.
Subsequent studies may aim to incorporate relevant methodological and theoretical insights from sampling approaches, primarily established in other fields (Sections \ref{subsection error-based stopping criteria} and \ref{subsection absolute error MC}).

\subsubsection{Stability-based criteria}
\label{subsubsection ALR stability}

The third class of stopping criteria relies on the stability of either the estimated failure probability or the limit state surface across enrichment iterations.
For instance, in \cite{Basudhar2010}, the relative change in successive polynomial coefficients of the SVM (Section \ref{subsection support vector machines}) is monitored, and the process is terminated once the change becomes sufficiently small.
However, such stabilization rules may risk premature convergence if the initial surrogate is highly inaccurate.
\textcolor{black}{This difficulty is particularly relevant in active learning reliability methods, where enrichment decisions are localized and may fail to fully capture complex failure topologies.
In problems involving multimodal limit state functions or highly nonlinear responses, the surrogate model may stabilize around a subset of failure regions while overlooking others, leading to biased yet apparently converged estimates.} 

In \cite{AN2024106878}, the relative change in consecutive minimum probable target points is employed as a stopping criterion in reliability-based design optimization (Section \ref{subsection reliability-based design optimization}).
Similarly, in \cite{gasper_teixeira_guedessoares_2015, WANG2022108214}, successive relative errors are used to determine termination.

Moreover, in adaptive Kriging high-dimensional reliability assessment method \cite{LIANG2025103827}, hybrid stopping criterion is proposed, combining a stability check on the convergence of failure probability estimates with an enhanced error-based stopping criterion.
In the AGP-SYS framework \cite{WEN2025103805}, an error- and stabilization-based stopping criterion is adopted, terminating the adaptive learning once ten successive training points yield prediction deviations below a fixed tolerance, thereby ensuring stable surrogate accuracy near the critical failure region.
Similarly, in the multi-mode failure boundary exploration and exploitation framework \cite{ZHANG2023103473}, stopping is governed by error- and stabilization-based rules, where surrogate refinement halts once the maximum failure probability error drops below a tolerance, and the final adaptive importance sampling stage terminates when both the coefficient of variation and the relative error of successive failure probability estimates satisfy prescribed thresholds.
In \cite{SU2025111059}, two stopping criteria are examined: one based on the stability of successive surrogate-based CDF estimates, and the other on the confidence bounds of the estimated distribution.

\color{black}
Stabilization-based criteria are effective primarily because they directly reflect the apparent convergence behavior of the surrogate model or failure probability estimate, providing an intuitive signal for termination.
However, as also discussed in the classical setting (Section \ref{subsection stability-based stopping criteria}), stability alone does not guarantee convergence to the true failure probability.
In particular, apparent stabilization may arise from incomplete exploration of the failure domain, especially in the presence of disconnected or rare failure regions, surrogate misclassification near complex or highly curved limit state boundaries, or local overfitting that reduces variability without improving global accuracy.

These limitations indicate that stability should be interpreted as a local convergence indicator rather than a sufficient measure of correctness.
In practice, this necessitates combining stability-based criteria with complementary indicators, such as error-based or probabilistic measures, to mitigate the risk of misleading convergence signals and to ensure more robust stopping behavior in complex structural reliability problems.

\color{black}
\subsubsection{Further related works}
\label{subsubsection further related works}

Recent developments have proposed more specialized stopping strategies tailored to complex problem settings.
For example, in active learning line sampling approaches (Section \ref{subsection line sampling}), such as SBALQ \cite{DANG2024110052}, PBAL-LS \cite{DANG2024116828}, and BAL-LS \cite{DANG2023102351}, stopping criteria are defined in terms of the stabilization of the acquisition function along lines or the convergence of failure probability estimates (see, also, \cite{DANG2023116068, DANG2024110053, SONG2021107113}).
In the combined Kriging-subset simulation \cite{CHEN2021107737}, a two-stage criterion is devised.
At the local level, it checks the relative model error ratio within each subset to confirm that the Kriging surrogate achieves sufficient accuracy, while at the global level, it constrains the accumulated error across subsets to ensure that the resulting probability estimate remains consistent and dependable.

For high-dimensional or multifidelity settings (Section \ref{subsection multifidelity models}), stopping conditions are informed by the collective contribution of multiple fidelity levels, with the adaptive learning strategy terminating once additional enrichment at lower or higher fidelity yields negligible improvement in prediction accuracy \cite{YIN2022102151, ZHANG2022102141}.
This type of strategy is particularly relevant when the computational cost associated with high-fidelity models is substantial \cite{yi_active-learning_2021}.
Similarly, in the AK-PSO-HHs framework \cite{THEDY2023103513}, the algorithm introduces a multi-layer stopping mechanism, involving the most probable point (MPP) and the $U$- and REIF functions (Section \ref{subsubsection learning functions}).
Also, stopping criteria combine thresholds on multiple learning functions \cite{YANG2024102426}, uncertainty quantification \cite{WANG2020101937}, and convergence of estimated statistics \cite{KIM2024102404}.
Stopping rules are adapted for dynamic reliability \cite{ALLAHVIRDIZADEH2023109406}, noisy data-driven models \cite{PIRES2025102543}, time- and space-dependent systems \cite{ZHAN2025110536}, and probabilistic fatigue life prediction under overloads \cite{JIANG2022103343}.

Another line of work considers the decomposition of the input space into subdomains, where localized surrogates are constructed and refined.
For instance, a sequential decomposition approach is proposed in \cite{GERASIMOV2023102364} to detect failure regions adaptively, with stopping criteria based on the stabilization of surrogate predictions and failure surface localization within subregions.
Similarly, the concept of stepwise uncertainty reduction, as used in optimal experimental design for fatigue life prediction \cite{COUSIN2024102483}, can be extended to reliability analysis by treating uncertainty reduction as a stopping indicator, particularly in high-dimensional settings where exploration of the entire domain is computationally prohibitive.
In \cite{ZHANG2023109034}, a stopping criterion is proposed for subdomain-based assessments, defined through the expected upper bound of the relative error in failure probability, where the target threshold is set by quantifying how the chosen value influences this relative error.

Lastly, it is worth repeating that stopping criteria employed for active learning reliability methods are generally empirical and tailored to specific problems, with little rigorous foundation \cite{TEIXEIRA2020106908}. 
In practice, heuristic approaches, such as threshold tuning, response smoothing, or monitoring iteration stability, are often adopted to avoid premature or excessively delayed termination.
Common practices include the requirement that convergence be maintained over successive iterations or the joint use of multiple stopping rules, though the latter may not necessarily enhance performance significantly \cite{MOUSTAPHA2022102174}.
Advancing explicit accuracy indicators and rigorously defined stopping principles stands out as a compelling direction for future study.

\subsection{Bayesian methods}
\label{subsection bayesian methods}

Bayesian methods have become increasingly important in structural reliability analysis, particularly for rare event estimation, as they quantify uncertainty and update probabilistic beliefs as data accumulate.
They provide a principled way to combine prior knowledge with observations through posterior inference.
Building on this foundation, Bayesian active learning has attracted growing attention by framing failure probability estimation within a Bayesian setting where the posterior distribution guides both sampling and stopping.
As new evaluations of the limit state are collected, the posterior mean and variance are updated sequentially, steering sample placement and convergence checks.

Most Bayesian active learning approaches rely on Gaussian process surrogates of the limit state, updated iteratively as new data are acquired.
Predictive uncertainty then drives both exploration of poorly understood regions and exploitation near the limit state.
Stopping is typically triggered when these functions fall below a tolerance, signaling diminishing returns from further sampling. More broadly, stopping criteria are embedded into the logic of Bayesian updating and often based on one or more of the following: decay of the learning function (Section \ref{subsubsection learning Bayesian}), error and stabilization of the failure probability estimate (Section \ref{subsubsection error stabilization Bayesian}), or reduction of posterior variance (Section \ref{subsubsection posterior Bayesian}).

We note that, while not inherently Bayesian, many active learning reliability methods (Section \ref{subsection active learning}) are implemented with Kriging models (Gaussian processes), which themselves are based on Bayesian inference.
In practice, many of those adopt Bayesian surrogates, making the boundary between the two categories quite blurred.
In a sense, Bayesian methods are broader and concern the explicit use of Bayes' theorem to update probability distributions as new information becomes available.
In structural reliability analysis, this covers Bayesian model updating, Bayesian networks, Bayesian MCMC/IS approaches, and Bayesian treatment of epistemic uncertainty.
Active learning based on Kriging can thus be thought of as a subset of this larger Bayesian perspective, while not all Bayesian approaches are active learning either.
For instance, Bayesian model updating or hierarchical modeling in reliability do not necessarily involve active learning.

\subsubsection{Learning functions}
\label{subsubsection learning Bayesian}

Similar to active learning reliability methods (Section \ref{subsection active learning}), Bayesian methods rely critically on a learning, or acquisition, function, which directs the active learning procedure by identifying the most informative data points.
Let us emphasize that the following list remains far from comprehensive, as even further classes of learning functions are likely to arise within the Bayesian framework, particularly driven by ongoing advances that foster novel acquisition strategies tailored to structural reliability analysis.

Hence, on the one hand, the $U$-function (Section \ref{subsubsection learning functions}) has also been employed in this context 
to adaptively select new training samples for the Kriging model (Section \ref{subsection kriging models}).
For instance, in the adaptive Kriging-based Bayesian updating strategy \cite{WANG2020101915}, its stopping condition is based on the expected variance of misclassification of samples in the accepted domain, using the $U$-function.
In a similar manner, the folded normal based expected improvement function (FENEIF) \cite{SHI2020106857} is employed in the context of Bayesian active learning \cite{Wang2023SSRN}.

On the other hand, a wide range of learning functions have been developed for Bayesian methods in structural reliability analysis.
For instance, in \cite{doi:10.1061/AJRUA6.0001179}, a learning function, termed upper bound posterior variance contribution (UPVC), is developed:
\[
 {\rm UPVC}({\bf x}):=\sqrt{\Phi\left(-\frac{\mathbb{E}_D[\widehat{g}({\bf x})]}{\sqrt{{\rm Var}_D(\widehat{g}({\bf x}))}}\right)\Phi\left(\frac{\mathbb{E}_D[\widehat{g}({\bf x})]}{\sqrt{{\rm Var}_D(\widehat{g}({\bf x}))}}\right)}f({\bf x}|\theta^{\star}),
\]
where $\widehat{g}({\bf x})$ denotes the Gaussian process prior at ${\bf x}$, $f(x|{\bm \theta}^{\star})$ is a joint probability density function with its parameter value $\theta^{\star}$ given, and $\mathbb{E}_D$ and ${\rm Var}_D$ denote the posterior expectation and variance operators conditionally on the observed data set.
As its variant, the penalized quasi posterior variance contribution (PQPVC) is proposed in \cite{DANG2024103613} with a penalty factor.
If the penalty factor is one, it
reduces to the quasi posterior standard deviation of the failure probability, as a special case of upper bound posterior variance contribution.
Introducing the penalty factor is essential, as it enables point selection through straightforward optimization of the PQPVC function under various penalty factors.

Moreover, in \cite{DANG2024110052}, a learning function is proposed in the form of 
\[
 \mathcal{J}_n({\bf u}):=\Phi\left(-\frac{m_{\mathcal{G}_n}({\bf u})}{\sigma_{\mathcal{G}_n}({\bf u})}\right)\prod_{k=1}^d \phi(u_k),
\]
where $\mathcal{G}_n$ hereafter represents the posterior distribution of $g(T^{-1}(U))$ (Section \ref{subsection preliminaries}) after $n$ observations are incorporated.
In this context, the point attaining the highest function value is typically regarded as the most promising candidate for selection.
In a similar manner, three learning functions, 
left-shifted contribution (LSC), right-shifted contribution (RSC) and left-shifted and right-shifted contribution (LSRSC) are developed in \cite{DANG2024116828}, for instance, the first one formulated as: 
\[
 {\rm LSC}({\bf u}):=\left[\Phi\left(-\frac{m_{\mathcal{G}_n}({\bf u})}{\sigma_{\mathcal{G}_n}({\bf u})}\right)-\Phi\left(-\frac{m_{\mathcal{G}_n}({\bf u})}{\sigma_{\mathcal{G}_n}({\bf u})}-b\right)\right]\prod_{k=1}^d \phi(u_k),
\]
measuring the contribution at the point ${\bf u}$ to the difference between the posterior mean and the left-shifted posterior mean of the failure probability.
Pseudo posterior variance contribution (PPVC) is given in \cite{HU2024102409}:
\[
 {\rm PPVC}({\bf u}):=\left[\Phi\left(-\frac{m_{\mathcal{G}_n}({\bf u})}{\sigma_{\mathcal{G}_n}({\bf u})}\right)\Phi\left(\frac{m_{\mathcal{G}_n}({\bf u})}{\sigma_{\mathcal{G}_n}({\bf u})}\right)\right]^{\alpha/2}\prod_{k=1}^d \phi(u_k),
\]
which measures the contribution of numerical uncertainty at point ${\bf u}$ to a pseudo posterior variance.
Moreover, weighted epistemic uncertainty contribution (WEUC) is proposed in \cite{DANG2025102539}:
\[
 {\rm WEUC}({\bf u}):=\sigma_{\mathcal{G}_n}({\bf u})\Phi\left(-\frac{m_{\mathcal{G}_n}({\bf u})}{\sigma_{\mathcal{G}_n}({\bf u})}\right)\Phi\left(\frac{m_{\mathcal{G}_n}({\bf u})}{\sigma_{\mathcal{G}_n}({\bf u})}\right)\prod_{k=1}^d \phi(u_k),
\]
where the first term $\sigma_{\mathcal{G}_n}({\bf u})$ puts a weight on the rest, referred to as the epistemic uncertainty contribution.
That is, the first term reward points with large uncertainty (hence to increase the exploration).
In particular, we refer the reader to \cite[Section 3 and Table 1]{DANG2025102539} for a general review on a variety of those related methods, including ALPI \cite{doi:10.1061/AJRUA6.0001179}, PBALC \cite{DANG2024116828}, SBALQ \cite{DANG2024110052}, PBPI \cite{HU2024102409}, and QBALC \cite{DANG2024103613}.


In the parallel adaptive-Bayesian failure probability learning (PA-BFPL) method \cite{DANG2022102259}, the expected misclassification
probability contribution (EMPC), defined as
\[
 {\rm EMPC}({\bf x}):=\Phi\left(-\frac{|m_{\mathcal{G}_n}({\bf x})|}{\sigma_{\mathcal{G}_n}({\bf x})}\right)\frac{f({\bf x})}{h({\bf x})},
\]
is employed, where $h({\bf x})$ is a suitable importance sampling density.
In the context of Bayesian active learning line sampling, posterior standard deviation contribution (PSDC) is proposed in \cite{DANG2023102351}:
\[
 {\rm PSDC}({\bf u}^{\perp}):=\phi_{\perp}({\bf u}^{\perp})\int k_{\widetilde{\Phi}_n(-\widetilde{\beta})}({\bf u}^{\perp},{\bf u}^{\perp '})\phi_{\perp'}({\bf u}^{\perp'})d{\bf u}^{\perp '},
\]
where $k_{\widetilde{\Phi}_n(-\widetilde{\beta})}$ denotes the posterior covariance functions of $\Phi(-\beta)$ after $n$ observations are given.
This learning function can be understood as a metric that reflects the influence of the epistemic uncertainty at the perpendicular component ${\bf u}^{\perp}$ (corresponding essentially to $(z_2,\cdots,z_d)$ in the formulation \eqref{LS formulation}) to the posterior variance of the failure probability (or, equivalently, the posterior standard deviation).
In a similar manner, upper-bound posterior standard deviation contribution (UPSDC) and approximate posterior standard deviation contribution (APSDC) are introduced, respectively, in \cite{DANG2023116068} and \cite{DANG2024110053}.
Finally, in \cite{DANG2025118092}, three established learning functions in Bayesian optimization (the lower confidence bound (LCB), probability of improvement (PI) and expected improvement (EI) \cite{jones_schonlau_welch_1998}) are specifically adapted to address time-variant reliability problems (see Section \ref{subsection time-variant reliability analysis}).

In a similar manner to active learning reliability methods (Section \ref{subsection active learning}), the choice of a learning function plays a central role in guiding the sequential selection of new samples, thereby controlling the balance between exploration (reducing global uncertainty) and exploitation (refining regions near the limit state surface).
While the $U$-function remains the most widely used due to its simplicity and direct link to the misclassification probability, more sophisticated functions 
can be effective by explicitly quantifying epistemic uncertainty in the surrogate model.
A pragmatic approach is to begin with a simple and interpretable metric like the $U$-function, and progressively adopt more advanced or hybrid functions as the surrogate matures, enabling a context-aware trade-off between reliability accuracy, sampling efficiency, and model confidence.

\subsubsection{Error- and stability-based stopping criteria}
\label{subsubsection error stabilization Bayesian}

A reliability updating framework is proposed in \cite{FENG2023102366}, combining Bayesian inference, sequential learning, and Kriging, with a stopping rule based on the relative error of the probability estimate.

A hierarchical stopping criterion is developed in \cite{Wang2023SSRN}, combining iteration-wise stabilization checks of surrogate predictions with a final error-bound rule from Chebyshev's inequality.
The process terminates once local accuracy and global probability estimation are sufficiently reliable, balancing efficiency and accuracy.

A semi-Bayesian active learning quadrature method is introduced in \cite{DANG2024110052}, where the stopping criterion is error- and stabilization-based, halting when the relative error between the posterior mean estimate and a penalized version falls below a threshold, with hyper-shell simulation approximating the required integrals.

In \cite{DANG2024116828}, the partially Bayesian active learning cubature (PBALC) methods propose three stopping rules as relative error checks between the posterior mean of the failure probability and its left-/right-shifted versions from credible bounds.
These ensure the process halts once the Bayesian predictor of the failure probability stabilizes within tolerance, enabling efficient estimation of very small probabilities.

The effectiveness of error- and stability-based stopping criteria holds owning to their reliance on model refinement guided by statistically meaningful indicators of convergence, such as the stabilization of posterior estimates or the decrease in relative prediction errors.
Criterion selection should reflect the inference context.
For instance, posterior mean- or variance-based rules suit continuous updates, while relative error or credible-bound checks are preferable for rare-event estimation.
Future research may pursue hierarchical criteria that adjust thresholds with posterior contraction, and incorporate diagnostics to guard against underfitting and overconfidence in surrogate predictions.
Also, as noted earlier in Sections \ref{subsubsection ALR error} and \ref{subsubsection ALR stability}, forthcoming studies may aim to integrate theoretical developments from other research fields (Sections \ref{subsection error-based stopping criteria} and \ref{subsection stability-based stopping criteria}).


\subsubsection{Posterior distribution-based stopping criteria}
\label{subsubsection posterior Bayesian}

Bayesian subset simulation (BSS) \cite{doi:10.1137/16M1078276} introduces a posterior distribution-based rule that halts evaluations once the expected misclassification probability falls below a prescribed fraction of the excursion probability, ensuring that further sampling meaningfully reduces posterior uncertainty.

In \cite{XIAO2022102230}, an adaptive importance sampling algorithm is proposed for Bayesian model updating, where the stopping criterion is defined by the normalized effective sample size (N-ESS).
Termination occurs once N-ESS exceeds a threshold, ensuring the constructed density adequately approximates the target posterior while balancing accuracy and cost.

A posterior distribution-based rule is proposed in \cite{FAROZ2023102354} that halts calibration once the Kullback-Leibler information gain stabilizes, indicating that the measurement error model is sufficiently informed.
This reduces calibration costs by limiting experiments to those that meaningfully decrease posterior uncertainty.

A quasi-Bayesian active learning cubature method is developed in \cite{DANG2024103613} with a posterior distribution-based criterion defined through the quasi-posterior coefficient of variation of the failure probability, enabling efficient assessment of extremely small probabilities.

In \cite{doi:10.1061/AJRUA6.0001179}, active learning augmented probabilistic integration (ALPI) employs a posterior distribution-based rule that stops once the posterior coefficient of variation of the failure probability falls below a threshold.
Using an analytically derived upper bound of the posterior variance, the method quantifies epistemic uncertainty efficiently.
Similar posterior coefficient of variation rules are adopted in partially Bayesian active learning line sampling (PBAL-LS) \cite{DANG2023116068}, parallel Bayesian probabilistic integration (PBPI) \cite{HU2024102409}, Bayesian active learning line sampling (BAL-LS) \cite{DANG2023102351}, and its log-normal variant BAL-LS-LP \cite{DANG2024110053}.

In \cite{DANG2025102539}, the weakly Bayesian active learning quadrature (WBALQ) method introduces a posterior distribution-based criterion using the first absolute central moment of the posterior failure probability as a tractable measure of epistemic uncertainty.
By halting once this measure is small relative to the posterior mean, the method achieves reliable estimation of very small probabilities without expensive variance evaluations.

In principle, posterior distribution-based stopping criteria achieve a balance between accuracy and computational cost by evaluating whether further sampling would substantially reduce posterior uncertainty.
The choice of criterion should align with the inference objective and the feasibility of the posterior measure adopted, such as normalized effective sample size in importance sampling (Section \ref{subsection importance sampling}) or posterior moment-based indicators in line sampling (Section \ref{subsection line sampling}).
A worthwhile research direction lies in consolidating these Bayesian stopping principles under a coherent decision-theoretic formulation, enabling adaptively evolving thresholds driven by posterior contraction and accommodating the nontrivial interplay between uncertainty and stopping rules in multifidelity and time-variant reliability analyses.

\subsection{Adaptive response surface methods}
\label{subsection adaptive response surface methods}

We next discuss stopping criteria for adaptive response surface methods, which build upon response surfaces (Section \ref{subsection response surfaces}) as surrogates of the limit state function.
In these methods, well-designed stopping rules are essential to balance predictive accuracy with computational efficiency by determining when further refinement of the surrogate is no longer necessary.
We note that the adaptivity here refers to sequentially updating the surrogate as additional sample points are incorporated, often guided by error estimates or reliability indices.
This differs from active learning reliability and Bayesian methods (Sections \ref{subsection active learning} and \ref{subsection bayesian methods}), which emphasize the targeted selection of new sample points in regions most informative for reliability estimation, for instance, in areas of high uncertainty or close to the limit state surface.

Early efforts in adaptive response surface methods focus on sequential refinement of polynomial approximations.
The sequential response surface method (SRSM) \cite{YINGWEILIU199439} iteratively shifts the expansion point closer to the design point when the previous surface proves insufficiently accurate, halting once error checks confirm convergence.
Similarly, in \cite{GUAN2001429}, a pragmatic approach is adopted, continuing refinement until reliability estimates stabilizes, rather than relying on formal statistical thresholds.
These early works illustrate how adaptivity is initially framed as stepwise updates driven by intermediate accuracy checks.

Later developments emphasize stability- and error-based stopping rules.
For instance, in \cite{HADIDI201715}, exponential response surfaces are refined until both the reliability index and limit state function at the design point converge within tolerance (Section \ref{section approximate methods}).
This principle is extended in \cite{LI201842} to radial basis function surrogates, where the enrichment halts once successive failure probability estimates differed by less than prescribed tolerances.
In relation, spline-based enrichment is terminated in \cite{ZHAO201769} once changes in the reliability index and failure probability became negligible, while polynomial and exponential surfaces are combined in \cite{10.1063/5.0074702}, halting updates once the reliability index stabilizes across iterations.
These approaches highlight a shift toward convergence criteria directly tied to reliability outputs.

Adaptive methods have also evolved to include sparse approximations and neural network surrogates (Section \ref{subsection neural networks}). 
In \cite{GUIMARAES201812}, an adaptive sparse polynomial response surface is proposed, where the procedure is stopped once the coefficient of variation of the estimated failure probability falls below a target threshold and cross-validation confirmed further enrichment unnecessary.
In \cite{GOMES2011186}, radial basis function neural networks are employed in reliability-based optimization, where stopping is dictated by genetic algorithm convergence rules such as reduced population diversity or reaching a maximum generation count. 
In the AK-ARBIS framework \cite{YUN2020101891}, surrogate refinement halts once the $U$-function criterion (Section \ref{subsubsection learning functions}) ensures correct classification of candidate samples and the coefficient of variation of the estimated failure probability drops below 5\%.
These examples illustrate how adaptivity is combined with different surrogate types, while retaining convergence rules focused on reliability accuracy.

Finally, spline-based adaptive frameworks further formalize stopping conditions using a posteriori error estimators.
In \cite{BRACCO2024117287}, hierarchical $C^1$ splines are employed over multi-patch domains, terminating refinement once error estimators show that additional enrichment would not meaningfully improve accuracy.
In parallel, the combination of multivariate adaptive regression splines (MARS) with successive response surface methodology (SRSM) \cite{4126283} relies on error-based stopping, halting when either relative errors fall below a threshold or differences between solutions in consecutive subregions become negligible.
Adaptive response surface methods such as ARSM  \cite{AltairFEKO_ARSM}, implemented in tools like Altair FEKO, refine the surrogate iteratively and may include a validation step against high-fidelity simulations to verify the estimated accuracy before terminating the adaptive process.
These works demonstrate how spline-based approaches integrate adaptive refinement with rigorous error estimators, further advancing the design of stopping criteria for response surface methods.

In essence, these stopping schemes succeed due to their foundation on explicit and data-informed convergence metrics that couple surrogate improvement with estimation enhancement.
Ongoing evaluation of the stabilization of key quantities, such as reliability indices, failure probabilities, and cross-validation errors, guarantees termination precisely at the point where further refinement ceases to be beneficial.
It would be of particular interest to design autonomous, uncertainty-aware stopping mechanisms that exploit probabilistic error quantification, multifidelity surrogates, and adaptive learning, ultimately enabling self-regulating adaptive response surface frameworks capable of addressing nontrivial problems.

\subsection{Reliability-based design optimization}
\label{subsection reliability-based design optimization}

In structural reliability analysis, reliability-based design optimization (RBDO) offers a systematic framework for determining optimal design configurations that satisfy prescribed reliability requirements, with its conceptual origin tracing back to \cite{RACKWITZ1978489} in the 1970s.
By embedding structural reliability evaluations into the optimization workflow, RBDO addresses uncertainties in input parameters, loading conditions, and system behavior.
A conventional RBDO setup typically employs a two-level iterative scheme, where reliability assessments guide the optimization at each step (see, for instance, \cite{choi2006reliability}).
Appropriate stopping criteria are thus essential for both loops to ensure convergence, accuracy, and computational efficiency.

In the outer loop, which optimizes design variables subject to reliability constraints, stopping measures include the maximum number of iterations \cite{app11104708, MISKA2025103755}, the relative error of the predicted response at selected sample points \cite{LEE20081463, LIEU2019599}, the standard convergence conditions of the gradient-based deterministic optimizer \cite{DUBOURG201347}, and changes in the objective function \cite{SONG2025120166, zhouRBDO}.
In a different vein, illustrative examples are presented in \cite{LI2019106432}, highlighting the performance contrast between cases of insufficient and sufficient data available.
In \cite{SHEN2022102153}, the stopping criterion applies to the outer loop of the quantile-based SORA iterations, where the inner reliability analysis loop is avoided by reformulating the constraints in terms of quantiles.
In \cite{WANG2025103793}, stopping is set based on the standard deviation of individual fitness values in the genetic algorithm population.

The inner loop evaluates the reliability of candidate designs, often using approximate methods such as FORM \cite{Lopez2012RBDO}. Various stopping criteria have been proposed, based on suitable quantities including the maximum number of iterations \cite{MISKA2025103755}, the relative error in successive augmented failure probability estimates \cite{SONG2025120166}, the spread between upper and lower surrogate-based bounds \cite{DUBOURG201347}, the intersection angle between predicted and true gradients of updating samples \cite{NAN2024109944}, the number of misclassified random points \cite{LI2022108539}, the relative change in consecutive minimum probable target points \cite{AN2024106878}, the misclassification error of cross-validated metamodels \cite{ALLAHVIRDIZADEH2023109406}, and the improvement in the reliability of active constraint functions after adding training points \cite{YANG2024102426}.
In the improved weighted average simulation method (WASM) \cite{OKASHA201647}, the inner loop stopping criterion is based on a convergence check, where the incremental probability of failure is terminated once the gap between its accumulated value and the computed upper bound falls below a specified tolerance.
In stochastic subset optimization \cite{TAFLANIDIS2008324}, the inner loop for subset identification imposes stopping criteria based on a ratio of probability contents and its coefficient of variation.
In \cite{WANG2025103793}, stopping is based on a variance-based convergence criterion, where active learning continues until the maximum predicted variance of the Kriging surrogate within the confidence domain falls below a specified threshold.
 
The strength of these stopping strategies lies in their use of quantitative measures that unify optimization progress with the precision of reliability estimation.
Continuous assessment of the stability of key quantities, such as objective values, gradients, probabilistic outputs, and surrogate uncertainties, ensures that both design and reliability processes cease exactly when further iterations offer diminishing gains.
Promising avenues for future research involve devising stopping mechanisms that harness uncertainty quantification and multifidelity modeling, thereby establishing more robust termination criteria for high-dimensional or time-variant RBDO problems.

\subsection{Time-variant reliability analysis} 
\label{subsection time-variant reliability analysis} 
Although time-variant reliability analysis has attracted sustained research interest \cite{Wang_2021, Zhang03032024}, the development of stopping criteria has, by comparison, received relatively little focus in previous studies. 
Most existing studies have concentrated on improving model accuracy or uncertainty quantification, while the systematic design of robust stopping rules has remained largely overlooked. 
Nevertheless, notable efforts have gradually emerged, proposing problem-specific convergence conditions that monitor quantities such as failure probability stabilization, predictive uncertainty reduction, or credible interval contraction. 

\textcolor{black}{As in the time-invariant case, the $U$-function is widely employed in appropriately modified forms (for instance, its time-dependent variant) to guide the stepwise improvement of the Kriging model, as illustrated in \cite{GUO2023103441, ZHANG2021113990, ZHUANG2024103598}.
In particular, it is adopted in \cite{ZHANG2021113990} to construct a Kriging model that approximates the most probable point (MPP) trajectory.
At each iteration, the algorithm selects the time instant associated with the smallest $U$-value, corresponding to the highest prediction uncertainty, and updates the surrogate model accordingly.
The process terminates once all $U$-values exceed 2, indicating sufficient prediction confidence across the time domain.
Because the $U$-function is evaluated over the temporal evolution of the MPP trajectory, the stopping criterion is intrinsically suited to time-dependent problems, ensuring reliable characterization of their dynamic behavior.
Notably, this selection mechanism implicitly follows a forward temporal progression, in which reliability information is accumulated causally from earlier to later times.}


A Bayesian credibility-based approach is introduced in \cite{DANG2025118092} that halts the learning process once the conservative and optimistic credible bounds of the time-dependent failure probability get close enough (as a time-variant version of a stopping criterion developed in \cite{DANG2024116828}), with an additional Monte Carlo-based precision check ensuring statistical reliability. 
In a related development, a single-loop framework is constructed in \cite{QIAN2021107261}, combining a multi-response Gaussian process with subset simulation. 
This framework selects new evaluation points using multiple learning functions (such as $U$, EFF, $H$, and EI (Section \ref{subsubsection learning functions})) and monitors the coefficient of variation of the estimated probabilities, thereby providing a quantitative criterion for learning saturation without resorting to nested optimization. 

The stopping criterion adopted in \cite{HU2021108029, hu_mahadevan_2016} is based on controlling the maximum relative error of the estimated failure probability. 
At each iteration, the algorithm evaluates the confidence of the Kriging predictions through the $U$-function, distinguishing accurately and inaccurately classified samples. 
The maximum possible relative error between the current and true failure probabilities is then computed, and the iterative learning process terminates once this error falls below 5\%. 
Because this condition directly monitors the accuracy of the time-dependent failure probability, it is inherently tailored to the time-variant nature of the reliability problem. 

In the adaptive spatiotemporal sampling physics-informed neural network (ASTS-PINN) \cite{HU2026111560}, the training process is stopped upon detecting stagnation of the loss function, specifically when the mean loss per epoch decreases by less than 1\% over 50 successive epochs. 
In \cite{REY2026111536}, a fatigue reliability analysis is carried out accounting for material variability in steel structures, where the training of the multifidelity Kriging model is terminated once fewer than $99\%$ of points in the domain satisfy the time-dependent learning function across all time steps. 
In the nonprobabilistic time-variant reliability framework developed in \cite{GONG2024103687}, sample enrichment is guided adaptively by the expected improvement (EI) function during the iterative construction of Kriging surrogate models for interval dynamic analysis. 
The iteration terminates when each new sample lies sufficiently close to existing ones in the parameter space. 

In the adaptive virtual model-assisted most probable point trajectory-based (AdaVM-MPPT) approach \cite{ZHAO2026111748}, a straightforward and practical stopping rule is adopted based on the temporal stabilization of estimated statistics rather than on strict theoretical convergence guarantees. 
Specifically, the simulation is terminated once the coefficient of variation or another time-progressive convergence indicator (such as the mean or variance of the estimated failure probability) remains nearly unchanged over several successive iterations. 

A time-dependent adaptive SVM-Monte Carlo (TASVM-MCS) strategy (Section \ref{subsection support vector machines}) is presented in \cite{ZHENG2025110510}, where the stopping rule directly monitors the relative variation in the time-variant failure probability function (TFPF) across times points in each iteration. 
By measuring convergence in the target reliability metric itself, rather than in surrogate-based indicators, this criterion ensures robust estimation of time-dependent failure probabilities. 

In the ensemble stochastic configuration network-based sequential stratified importance sampling (eSCN-SSIS) method \cite{LU2025103861}, convergence is controlled through two relative-error thresholds that evolve with time as the time-dependent failure probability is updated. 
The first threshold governs the iterative refinement of the eSCN-driven importance sampling density, halting the adaptive update once the relative change between successive estimates becomes sufficiently small. 
The second threshold, which is stricter, serves to verify that the estimated probability has stabilized after the sampling density reaches convergence, ensuring the robustness of the final time-variant prediction. 

In the integrated nested Laplace approximation (INLA) framework \cite{ZHOU2026111461}, its stopping criterion governs both the inner and outer iterative loops, where convergence is declared when the difference between consecutive parameter estimates falls below predefined tolerances, ensuring stable updates of the time-evolving latent parameters in the step-stress Wiener process throughout the time-variant degradation analysis. 

The recursive information trace optimization (RITO) \cite{ZHANG2026111795} addresses a time-variant setting in accelerated life testing (ALT), where reliability evolves dynamically under time-dependent stress. 
It adopts an information-based adaptive stopping rule combining precision monitoring (CI-width threshold) with recursive Fisher-information maximization, achieving a balanced trade-off between statistical efficiency and experimental cost. 

\color{black}
Note that most of the above stopping criteria implicitly adopt a causal viewpoint, in which convergence is assessed progressively along the forward time axis. This reflects a practical algorithmic design choice rather than a fundamental requirement of the underlying probabilistic models. In particular, theoretical constructs such as Girsanov's transformation suggest a form of time-reversal symmetry under measure equivalence, indicating that forward-only monitoring may not fully exploit the structure of time-variant problems.

From this perspective, existing stopping rules may overlook opportunities for computational reduction by not leveraging temporal symmetry or bidirectional information flow. While current methods emphasize stability of time-evolving reliability estimates, future developments could explore non-causal or symmetry-aware stopping strategies that utilize information from both past and future time instances, potentially improving efficiency without compromising accuracy.

More broadly, time-variant reliability analysis proceeds as a sequential process in which convergence must be assessed across both iteration and temporal evolution. The stopping criteria discussed above are effective insofar as they detect when failure probabilities and predictive uncertainty stabilize over time, thereby ensuring that computations continue only while they provide meaningful gains in accuracy. At the same time, their performance depends on how well this temporal evolution is captured and interpreted within the algorithmic framework.

Against this background, further developments may benefit from revisiting stopping criteria through the lens of temporal structure and measure equivalence, particularly in settings involving trajectory-based formulations and stochastic process representations.
Connections to approaches based on trajectory splitting \cite{CHING20051557}, Girsanov's transformation \cite{SUNDAR201332}, and probability density evolution \cite{doi:10.1061/(ASCE)EM.1943-7889.0001351} may provide a pathway toward more unified and efficient stopping strategies for time-variant reliability analysis.


\color{black}
\section{Synthesis, challenges, and outlook}
\label{section concluding remarks}

The motivation for the present article arises from the clear absence of comprehensive treatments of stopping criteria in the existing literature.
This deficiency mainly originates from the tendency for stopping rules to be developed alongside specific numerical methods, each characterized by features tailored to particular problem formulations.
While a few broadly applicable stopping criteria have been proposed in the literature, those rules often do not perform universally well in practice, aptly encapsulated by the {\it ``no free-lunch principle''} \cite{MOUSTAPHA2022102174} and the remark that {\it ``generality is usually inversely proportional to efficiency''} \cite{SCHUELLER2007167}.
Consequently, as highlighted throughout, the literature offers a diverse but fragmented array of stopping criteria, each designed to address the specific demands of various algorithms and computational frameworks.

Building upon this recognition, the present article aims to consolidate and systematize existing stopping criteria and related methodologies, providing a comprehensive reference for researchers and practitioners.
For the sake of completeness, key background concepts are revisited to ensure clarity and coherence, while the scope is intentionally limited to methodological synthesis, excluding detailed technicalities, performance assessments or benchmark comparisons of specific criteria.

To achieve this goal, we first develop in Section \ref{section introdution} a unified conceptual and methodological framework for the review, encompassing the taxonomy of stopping criteria (Section \ref{subsection taxonomy of stopping criteria}), their performance dimensions (Section \ref{subsection performance dimensions}), and the principles guiding their systematic organization (Section \ref{subsection review methodology}).
On this basis, and after introducing the structure of the article (Section \ref{subsection structure of the present overview}) and the necessary notation and technical preliminaries (Section \ref{subsection preliminaries}), we proceed to examine stopping criteria for approximate methods (Section \ref{section approximate methods}) and Monte Carlo simulation (Section \ref{section stopping criteria for monte carlo simulation}).

As for Monte Carlo simulation, starting with the basic principles (Section \ref{subsection basics of monte carlo simulation}), we reviewed criteria based on the coefficient of variation (Section \ref{subsection CoV}), the principled Monte Carlo simulation approach (Section \ref{subsection error-based stopping criteria}), and absolute-error-based formulations (Section \ref{subsection absolute error MC}), the latter discussed in the context of Chebyshev's inequality (Section \ref{subsubsection chebyshev}), the central limit theorem (Section \ref{subsubsection CLT}), Hoeffding's inequality (Section \ref{subsubsection hoeffding}), and stability-based criteria (Section \ref{subsection stability-based stopping criteria}).
These developments, though originating largely outside structural reliability, offer a rich foundation of concepts and tools that can be further adapted to advance sampling strategies, surrogate refinement, and convergence assessment in structural reliability analysis.
We then explored stopping criteria for advanced sampling techniques in structural reliability analysis (Section \ref{section sampling methods}), covering importance sampling, subset simulation, line sampling, directional sampling and Markov chain Monte Carlo, as presented in Sections \ref{subsection importance sampling}-\ref{subsection MCMC}.

Next, we present a high-level overview of stopping criteria across modeling and methodological frameworks in Sections \ref{section advanced modeling frameworks} and \ref{section advanced methodological frameworks}, emphasizing shared themes and trends rather than technical specifics.
First, in Section \ref{section advanced modeling frameworks}, we overviewed stopping criteria by modeling frameworks, including Kriging models,
response surfaces,
polynomial chaos expansions,
support vector machines,
neural networks,
multifidelity models,
reduced-order models, 
and ensemble models, 
as presented in Sections \ref{subsection kriging models}-\ref{subsection ensemble models}.
Then, in Section \ref{section advanced methodological frameworks}, we reorganized the discussion by advanced methodological approaches for structural reliability analysis, encompassing active learning reliability methods,
Bayesian methods,
adaptive response surface methods,
reliability-based design optimization, and
time-variant reliability analysis,
as presented in Sections \ref{subsection active learning}-\ref{subsection time-variant reliability analysis}.

\color{black}
\subsection{Synthesis and critical reflections}
\label{subsection synthesis and critical reflections}

Having examined a broad range of stopping criteria across numerical, modeling, and methodological frameworks, several overarching observations emerge. Despite the increasing sophistication of individual approaches, the field remains structurally fragmented, with many stopping criteria developed in close association with specific algorithms and problem settings. As a result, while numerous formulations exhibit strong empirical performance, their applicability beyond their original context is often limited, and a unified understanding remains elusive.
Broadly speaking, three central observations arise from this synthesis, as follows.

First, stopping criteria with rigorous theoretical guarantees remain exceptionally rare, like in simulated annealing \cite{vanLaarhovenAarts1987} and, more recently, in the training dynamics of deep learning \cite{GoodfellowBengioCourville2016}.
Existing theoretical results, particularly for error-based and stability-based criteria (Sections \ref{subsection error-based stopping criteria} and \ref{subsection stability-based stopping criteria}), typically rely on idealized assumptions such as independent and identically distributed sampling.
Even under such assumptions, stopping rules may malfunction or fail to detect convergence reliably \cite{malfunctioning, JWsurvey}.
Once these assumptions are relaxed, as is common in structural reliability analysis, the theoretical analysis becomes substantially more involved and in many cases analytically intractable, as already observed in related areas such as steady-state simulation and martingale-based estimation \cite{RK18, doi:10.1287/mnsc.28.5.550, survey_SteadyStateI, JWmartingale}. 
These difficulties highlight a persistent gap between theoretical guarantees and practical applicability, particularly in complex, dependent, or rare-event settings.

Second, across much of the literature, stopping criteria remain predominantly empirical and problem-specific.
In many frameworks, particularly active learning reliability methods (Section \ref{subsection active learning}), termination rules are defined through heuristic thresholds, smoothing strategies, or stability checks applied over successive iterations \cite{GASPAR201424, TEIXEIRA2020106908}.
While such approaches are often effective in practice, they require careful tuning and may exhibit sensitivity to model behavior, sampling strategy, or problem geometry.
This empirical character is not incidental but reflects deeper structural challenges. For instance, in subset simulation (Section \ref{subsection subset simulation}), convergence depends on adaptive thresholding and Markov chain behavior \cite{AU2007183, KATAFYGIOTIS2007194}, whereas in line sampling (Section \ref{subsection line sampling}), performance is strongly influenced by the geometry of the failure domain \cite{Pradlwarter_2007}. Similar limitations persist in Bayesian formulations (Section \ref{subsection bayesian methods}), where stopping thresholds remain user-defined and key quantities, such as posterior variance, are difficult to evaluate analytically \cite{doi:10.1137/16M1078276, doi:10.1137/S0097539797315306, DANG2023102351}.
Overall, these observations suggest that stopping behavior is primarily governed by empirical adequacy rather than first-principles derivation, with parameters such as coefficient of variation or tolerance thresholds serving as de facto indicators calibrated to balance accuracy and computational cost.

Third, recent developments indicate a clear shift from isolated stopping rules toward hybrid and integrative strategies.
Rather than relying on a single criterion, many modern approaches combine multiple indicators, such as error-based, variance-based, and stability-based measures, to improve robustness and generality.
For example, multiple learning functions are jointly employed to enhance convergence in active learning frameworks \cite{DANG2025118092, YANG2024102426}, while hybrid formulations that integrate stability checks with refined error-based stopping have been proposed to balance statistical rigor and computational efficiency \cite{LIANG2025103827}.
These developments reflect a broader transition from single-rule heuristics to coordinated, multi-indicator strategies.
At the same time, achieving effective integration requires a deeper understanding of the often competing roles of accuracy, stability, and efficiency, which remain only partially understood \cite{MOUSTAPHA2022102174}.

Taken together, these observations reveal a fundamental structural tension in the field.
While stopping criteria are widely used and often empirically successful, their behavior is insufficiently understood from a unified probabilistic and statistical standpoint.
This tension motivates the need for clearer conceptual distinctions, more structured comparisons, and synthesis-oriented frameworks that can guide both interpretation and practical application.

\color{black}

\subsection{Prospects for future research}
\label{subsection prospects for future research}

Setting aside, for the moment, the fundamental questions of theoretical justification and broad applicability (Section \ref{subsection synthesis and critical reflections}), the exploration of stopping criteria as a whole reveals a broad and fertile landscape for research in structural reliability analysis.
Further opportunities may arise from integrating stopping rules with machine-learning surrogates, multifidelity modeling, and physics-informed neural networks, which together can enhance accuracy and efficiency in contexts where conventional methods often struggle.
In particular, extending these developments to time-variant reliability analysis represents an important frontier for achieving adaptive and temporally consistent termination strategies.
At the same time, there is growing value in formulating sequential and adaptive criteria that jointly account for model uncertainty, sampling variability, and computational cost, enabling more balanced, data-driven, and automatic stopping decisions.
Equally important is the creation of frameworks that remain dependable under epistemic uncertainty, rare-event regimes, or changing system dynamics, where conventional stopping criteria tend to be unreliable \cite{malfunctioning}.

Let us turn to several specific topics that offer especially promising directions for future research.
First, parallel computing holds particular potential, moving beyond simple task-level splitting toward adaptive systems that flexibly distribute computational effort across simulation, surrogate learning, and reliability estimation, as demonstrated in recent contributions (for instance, \cite{LI2023109513, ZHAO2024103573, ZHENG2026111493, ZHOU2025117524}). 
The growing use of multi-core CPUs, GPUs, and cloud platforms may make it possible to exploit concurrency at many scales. 
It would be of relevance to further investigate asynchronous sampling, GPU-based subset simulation, and distributed active learning, along with scalable variance estimation and reliable convergence checks, so that large-scale and time-variant analyses (Section \ref{subsection time-variant reliability analysis}) can fully benefit from high-performance computing.

Another important direction is the integration of stopping criteria with multifidelity approaches (Section \ref{subsection multifidelity models}).
Developing cross-level stopping rules that allocate resources based on variance or information gain could help balance computations between coarse and detailed models.
Establishing theoretical convergence results for such hybrid frameworks, which link active learning (Section \ref{subsection active learning}), variance reduction (Section \ref{section sampling methods}), and fidelity adaptation, would be crucial for creating self-adaptive and efficient workflows.

Addressing dimensionality challenges remains a key issue.
When many uncertain parameters interact, the convergence behavior becomes harder to assess \cite{SCHUELLER2007167}.
For instance, it may merit attention to design dimension-adaptive stopping rules that test convergence in reduced or active subspaces identified through sensitivity or manifold learning.
Combined with sparse surrogates and low-rank approximations, these approaches may yield scalable, dimension-robust convergence metrics and self-adaptive algorithms suitable for complex system-level reliability analyses.

It is also essential to better understand the profound link between convergence in traditional simulations and active learning.
Conventional approaches rely on statistical indicators such as the coefficient of variation, while active learning relies on measures of surrogate-model uncertainty.
Coupling simulation variance with learning uncertainty to dynamically balance exploration and convergence represents a promising research direction, offering a pathway toward self-regulating algorithms capable of achieving robust and non-redundant termination behavior, especially in high-dimensional or time-variant problems (Section \ref{subsection time-variant reliability analysis}).

Beyond these technical advances and investigations, the discipline would benefit from renewed benchmark initiatives and rigorously defined performance metrics, well exemplified by the comprehensive benchmarking study on active learning \cite{MOUSTAPHA2022102174} and even the dedicated journal issue in the last decade \cite{StructSaf2007SpecialIssue}, this time alongside a greater emphasis on the comparative evaluation of stopping criteria.
The advancement of shared benchmarks, openly documented case studies, and clearly defined performance metrics for comparing stopping criteria would provide an essential foundation for objective evaluation and reproducibility.
A particularly promising step in this direction would be to enhance the well-established repository \cite{RPrepo2019}, for instance, by incorporating capacities for multifidelity (Section \ref{subsection multifidelity models}) and time-variant reliability analysis (Section \ref{subsection time-variant reliability analysis}).
Such resources would not only facilitate fair comparison among competing methods but also accelerate their translation into engineering practice by clarifying their strengths, limitations, and domains of applicability.





The research directions outlined here represent only a fraction of the vast and evolving research landscape, each meriting systematic, sustained, and interdisciplinary exploration.
We hope that this article will not only serve as a reference point but also as a catalyst for the continued refinement, integration, and theoretical consolidation of stopping criteria, ultimately contributing to more robust, transparent, and reliable estimation of failure probabilities in structural reliability analysis.





\color{black}


\bibliographystyle{abbrv}
\bibliography{ref_stopping}












\end{document}